\pdfoutput=1
\documentclass[letterpaper]{article} %

\usepackage[preprint]{aaai2027}  %
\usepackage[hyphens]{url}  %
\usepackage{graphicx} %
\usepackage{natbib}  %
\usepackage{caption} %
\usepackage{algorithm}
\usepackage{algorithmic}

\usepackage{newfloat}
\usepackage{listings}
\DeclareCaptionStyle{ruled}{labelfont=normalfont,labelsep=colon,strut=off} %
\floatstyle{ruled}
\newfloat{listing}{tb}{lst}{}
\floatname{listing}{Listing}

\usepackage{booktabs}

\usepackage{amsmath,amssymb,amsthm}
\usepackage{cite}
\usepackage{tikz}
\usepackage{todonotes}
\usepackage[T1]{fontenc}
\usepackage[utf8]{inputenc}
\usepackage{textcomp} 
\usepackage{comment}
\usepackage{booktabs}
\usepackage{multirow}
\usepackage{graphicx}
\usetikzlibrary{automata, arrows.meta, positioning}

\newtheorem{definition}{Definition}
\newtheorem{lemma}{Lemma}
\newtheorem{theorem}{Theorem}
\newtheorem{corollary}{Corollary}
\newtheorem{remark}{Remark}
\usepackage{lineno}

\usepackage{amsmath,amssymb,amsthm}   %
   \newcommand{\A}{\mathcal{A}}          %
   \newcommand{\C}{\mathcal{C}}          %

\title{Certifying Almost-Sure Reachability in Unknown MDPs}
\author{
    S. Akshay\textsuperscript{\rm 1},
    Suguman Bansal\textsuperscript{\rm 2},
    Gaurav Mahla\textsuperscript{\rm 1}
}
\affiliations{
    \textsuperscript{\rm 1}Department of Computer Science and Engineering,
    Indian Institute of Technology Bombay, Mumbai, India\\
    \textsuperscript{\rm 2}School of Computer Science,
    Georgia Institute of Technology, Atlanta, GA, USA\\
    akshayss@cse.iitb.ac.in,
    suguman@gatech.edu,
    mgauravmahla@gmail.com
}

\newcommand{\M}{\mathcal{M}}

\begin{document}

\maketitle

\begin{abstract}
{\em Specification-Guided Reinforcement Learning (RL)} has largely concerned itself with \emph{learning}
an optimal policy against a formal specification in an unknown {\em Markov Decision
Process (MDP)}.
We take up the dual question. Given a policy
already in hand, can one \emph{verify} that it is correct against a formal
specification in an unknown MDP, using sampling access alone? Such a question arises
naturally in safety-critical systems. We investigate it for the {\em almost-sure
reachability property}, whose correctness is witnessed by a {\em ranking
supermartingale (RSM)} certificate. Our first finding is negative. In the absence of
any further information about the MDP, one can neither learn such a certificate nor
verify a candidate one with probabilistic guarantees. %
On the positive side, we show that, under a mild restriction, a known lower bound
$p_{\min}$ on nonzero transition probabilities, a correct certificate \emph{can} be learned with PAC-style guarantees. To the best of our knowledge, this is the first study of learning and verifying certificates for almost-sure reachability in
unknown MDPs from sampling access alone.
Finally, our experiments show that the $p_{\min}$-based sufficient condition, while sound, is conservative: the learned certificate often remains valid on the true MDP even when the condition fails.
\end{abstract}

\section{Introduction}
\label{sec:intro}

Reinforcement learning (RL)~\cite{suttonbarto2018rl} studies how an agent should act
in a stochastic environment, standardly modeled as a \emph{Markov decision process}
(MDP)~\cite{puterman1994markov}. In many deployments the transition probabilities are
not known explicitly. The resulting \emph{unknown MDPs} grant the agent only
\emph{sampling access}. Querying a state--action pair $(s,a)$ returns a next state
$s'\sim P(\cdot\mid s,a)$, while the kernel $P$ itself stays hidden. 
The dominant
question in this setting, particularly when considering specification-guided RL~\cite{alur2026specification},
has been one of \emph{synthesis}, namely how to \emph{learn}
a policy that is optimal against a given specification.

We take up a dual and equally pressing question. Given a policy already in hand, can
we \emph{verify} that it is correct against a formal specification, again using only
sampling access to the unknown MDP? A positive answer would let us certify a deployed
controller without ever building a model of its environment. A negative one would
tell us precisely what extra knowledge such certification demands. It is especially important in safety-critical domains such as robotics and
autonomous systems. In effect, we initiate the study of \emph{certificate learnability}, the question of whether a proof of correctness can be learned or verified from samples, as the counterpart to the well-studied \emph{policy learnability}. Throughout, we fix the policy $\pi$ and work with unknown finite MDPs.

Verification of this kind cannot rest on simulation alone. Even in a finite MDP,
executions may be arbitrarily long and infinitely many, so success on sampled runs
cannot rule out an execution that violates the specification forever. What is needed
instead is a \emph{certificate}, a mathematical witness that a policy satisfies its
specification without enumerating all executions. Our specification of interest is
\emph{almost-sure reachability}, which asks that from the initial state the policy
$\pi$ reach a target set $T\subseteq S$ with probability $1$. For this property,
\emph{ranking supermartingales} (RSMs)~\cite{lechner2022stability} are a natural
certificate. An RSM assigns each state a nonnegative rank that, outside the target,
strictly decreases in expectation after one step. The process therefore makes
expected progress toward $T$ and, since the rank is nonnegative, cannot avoid $T$
forever. For finite MDPs an RSM is both a \emph{sound} and \emph{complete} witness to
almost-sure reachability.

Certifying almost-sure reachability thus reduces to producing, or checking, an RSM.
When $P$ is known this is straightforward, as the expected-decrease inequality is
directly verifiable. Under sampling access it is not. The decrease is an expectation
over the hidden kernel, so a candidate can only be estimated against a \emph{learned}
model $M'$ and its verdict transferred back to the true MDP $M$. It is not obvious
that a certificate can be produced, or even checked, from samples. We study three
questions together, along with the transferability that bridges them.
\emph{Learning} asks whether a correct RSM certificate can be produced from samples
with high probability. \emph{Verification} asks whether a candidate RSM can be checked
from samples with high probability. \emph{Transferability} asks when a certificate
valid on the learned $M'$ remains valid on $M$.
This differs from classical PAC-MDP \emph{policy} learning. For instance,
\citet{yang2021intractability,alur2021framework} show that PAC learnability for LTL
objectives holds exactly for formulas decidable within a finite horizon. Here the
policy is fixed and the object to be certified is the RSM itself.

Our first result is negative, and its cause is intrinsic to sampling-based
certification. A certificate cannot be produced or confirmed until the transitions
that make progress toward the target have been \emph{witnessed}, yet a progress
transition may carry arbitrarily small probability. Such a \emph{rare transition} can
remain unobserved for arbitrarily long, so the samples needed to observe it
cannot be bounded by the size of the problem alone. We show this \emph{rare-transition
obstacle} is decisive. In the absence of any further information about the MDP, neither
learning a correct certificate nor verifying a candidate one is possible with
probabilistic guarantees.

The impossibility is driven entirely by transitions of vanishing probability, and a
single mild and standard assumption removes it. A known lower bound
$p_{\min}$ on every nonzero transition probability~\cite{ashok2019pac} keeps a progress
transition from hiding behind an arbitrarily small probability, so progress toward the
target becomes observable in finitely many samples. Our positive result then follows a
\emph{model-based} route. Rather than learn an RSM directly, we learn a sufficiently
accurate model $M'$ (which standard finite-MDP model-learning provides from sampling
access) and compute on it the \emph{expected hitting-time} function
$V(s)=\mathbb{E}^{\pi,M'}_s[H_T]$, a natural RSM candidate that decreases by exactly one
in expectation. Because the expected-decrease condition depends on transition
probabilities, a certificate valid under $P'$ need not be valid under $P$. We prove
that a close enough approximation preserves the RSM property, and under $p_{\min}$ we
obtain an explicit threshold on the approximation error. Learning a transferable
certificate thereby reduces to learning a sufficiently accurate model, and a correct
RSM certificate can be learned with probabilistic guarantees.

\paragraph{Our Results.} Our key contributions are:
\begin{enumerate}
    \item \textbf{Impossibility of learning and verification.} For general unknown
    finite MDPs, neither learning a correct RSM certificate nor verifying a candidate
    one is possible with probabilistic guarantees from sampling access alone, due to
    rare transitions (Section~\ref{sec:pac_imp}).
    \item \textbf{Sampling bounds under $p_{\min}$.} A known lower bound $p_{\min}$ on
    nonzero transition probabilities makes progress toward the target observable with
    high probability using finitely many samples, removing the rare-transition
    obstruction (Section~\ref{subsec:p_min-sampling}).
    \item \textbf{Transferable expected-hitting-time certificates.} An RSM valid for a
    learned approximation $M'$ remains valid for the true MDP $M$ once the
    approximation error is small enough. We instantiate this for the expected
    hitting-time certificate, yielding an end-to-end learning guarantee with an
    explicit sample complexity (Section~\ref{sec:exp_hit}).
    \item \textbf{Empirical evaluation.} On a range of finite MDPs the transfer
    guarantee is sound but conservative, remaining valid on the true MDP even where
    the sufficient condition fails (Section~\ref{sec:experiments}).
\end{enumerate}

\paragraph{Related Work.}
Supermartingale certificates, rooted in stochastic Lyapunov
theory~\cite{kushner2014partial} and barrier certificates~\cite{prajna2007framework},
are the standard witness for temporal properties of probabilistic systems. Ranking
supermartingales certify almost-sure reachability and stability~\cite{lechner2022stability}
and extend to reach-avoid~\cite{zikelic2023learning}, safety and distributional
objectives~\cite{akshay2023distributional,akshay2024certified}, quantitative
bounds~\cite{abate2025qsc,abate2026value}, and probabilistic-program
invariants~\cite{batz2023probabilistic}, with Farkas certificates playing the analogous
role for reachability constraints~\cite{funke2020farkas}. A parallel line \emph{learns}
such certificates, representing them as neural networks and alternating a proposer with a
counterexample-producing verifier~\cite{ravanbakhsh2017counterexample,chatterjee2023learner,DBLP:conf/concur/AbateEGPR23,zikelic2023compositional}.
All of this work assumes a \emph{known} model, so the certificate conditions are concrete
constraints to solve and to check. We instead have only sampling access to an unknown
kernel, where those constraints can be estimated but not evaluated exactly, and our
impossibility result and transfer condition delimit precisely this gap.

For unknown MDPs, PAC reinforcement learning bounds the samples needed to learn a
near-optimal policy or an accurate model~\cite{brafman2001rmax,azar2013minimax}. General
LTL objectives are not PAC learnable~\cite{yang2021intractability,alur2021framework}, and
the guarantees that survive rest on a known lower bound $p_{\min}$ on nonzero transition
probabilities~\cite{ashok2019pac,DBLP:conf/aaai/PerezS024,svoboda2024rl}, the same assumption our
positive result needs and for the same reason. Weaker asymptotic guarantees are also
attainable~\cite{palasamudram2026rl,le2024rl}. All of these results learn a \emph{policy}.
We instead fix the policy and ask whether its correctness \emph{certificate} can be
learned or verified from samples.

\section{Preliminaries}
\label{sec:prelim}

A \emph{Markov decision process} (MDP) is a tuple $M = (S, A, P, s_0)$ with finite
state space $S$, finite action space $A$, transition kernel
$P : S \times A \to \mathcal{P}(S)$, and initial state $s_0 \in S$, where
$P(s' \mid s, a)$ is the probability of moving from $s$ to $s'$ under $a$. A
(deterministic, memoryless) \emph{policy} $\pi : S \to A$ fixes an action at each
state, collapsing $M$ into a Markov chain over $S$ whose \emph{trajectories} are
sequences $\tau = (s_0, s_1, \ldots)$ with $s_{k+1} \sim P(\cdot \mid s_k, \pi(s_k))$.
For a target set $T \subseteq S$, $\mathrm{Reach}(T)$ denotes the event that a
trajectory visits $T$, and $\mathbb{P}^{\pi}_{s}$ the induced measure over
trajectories started at $s \in S$.
Throughout, we work with \emph{unknown} MDPs, where $P$ is not given and the learner
has only \emph{sampling access}. Choosing $a_k$ at $s_k$ returns a successor
$s_{k+1} \sim P(\cdot \mid s_k, a_k)$, while the probabilities stay hidden. This is
the only mode of access we assume, and all our sample-complexity bounds are stated
against it.

A fixed policy makes $M$ evolve stochastically. A \emph{filtration}
$\mathcal{F}_0 \subseteq \mathcal{F}_1 \subseteq \cdots$ is a nondecreasing sequence of
$\sigma$-algebras, with $\mathcal{F}_k$ the information available up to time $k$, in our
setting the trajectory prefix $(s_0, \ldots, s_k)$. The one process we need is a
\emph{supermartingale}, whose conditional expectation only decreases over time, the
monotonicity our certificates exploit.
\begin{definition}[Supermartingale~\cite{chatterjee2023learner}]
\label{def:supermartingales}
A process $\{X_k\}_{k \geq 0}$ adapted to a filtration
$\{\mathcal{F}_k\}_{k \geq 0}$ is a \emph{supermartingale} if
$\mathbb{E}[X_{k+1} \mid \mathcal{F}_k] \leq X_k$ for all $k \geq 0$.
\end{definition}
We rely on one classical fact. A nonnegative supermartingale is unlikely to rise far
above its starting value.
\begin{theorem}
\label{thm:supermartingale_inequality}
Let $\{X_k\}_{k \geq 0}$ be a nonnegative supermartingale. Then for every
$\lambda > 0$,
$\;\mathbb{P}\bigl[\sup_{k \geq 0} X_k \geq \lambda\bigr]
  \leq \mathbb{E}[X_0] / \lambda.$
\end{theorem}

Our specification of interest is \emph{almost-sure reachability}, that from $s_0$ the
policy $\pi$ reaches a target set $T \subseteq S$ with probability $1$. To certify it
we use a \emph{ranking supermartingale} (RSM)~\cite{lechner2022stability}, a nonnegative
progress measure that strictly decreases in expectation outside the target.
Intuitively, if every non-target state moves closer to the target in expectation and
closeness cannot go negative, $T$ cannot be avoided forever. Introduced for general
stochastic systems, RSMs here act on the Markov chain $\pi$ induces on $M$, with the
general formulation deferred to sup. mat.~\ref{app:rsm_general}.
\begin{definition}[RSM for an MDP]
\label{def:rsm_for_MDP}
Let $M = (S, A, P, s_0)$ be a finite MDP, $\pi$ a fixed policy, and
$T \subseteq S$ a target set. A function $V : S \to [0, \infty)$ is an
\emph{$\varepsilon$-RSM} for $(M, \pi)$ w.r.t.\ $T$ if there is a constant
$\varepsilon > 0$ such that
\begin{itemize}
    \item[(i)] $V(s) \geq 0$ for all $s \in S$, and
    \item[(ii)] $V(s) \geq \mathbb{E}_{s' \sim P(\cdot \mid s, \pi(s))}[V(s')] + \varepsilon$ for
all $s \in S \setminus T$.
\end{itemize}
\end{definition}
Here $V(s)$ is a \emph{rank} measuring distance to the target. Condition (ii) forces
expected progress of at least $\varepsilon$ at every non-target state, and since $V$ is
nonnegative this cannot continue indefinitely, so $T$ must be reached. More precisely,
\begin{theorem}
\label{thm:probability_one_if_rsm_exists}
Let $M = (S, A, P, s_0)$ be a finite MDP, $\pi : S \to A$ a fixed policy, and
$T \subseteq S$ a target set. If an $\varepsilon$-RSM
$V : S \to [0, \infty)$ for $(M, \pi)$ w.r.t.\ $T$ exists, then
$\mathbb{P}^{\pi}_{s}[\mathrm{Reach}(T)] = 1$ for every $s \in S$; in particular
$\mathbb{P}^{\pi}_{s_0}[\mathrm{Reach}(T)] = 1$.
\end{theorem}

The proof (deferred to sup. mat.~\ref{app:omitted}) builds a nonnegative supermartingale that would grow without bound if the
target were missed, then applies Theorem~\ref{thm:supermartingale_inequality} to rule it
out. %
RSMs are thus \emph{sound}, and in finite MDPs also \emph{complete}. An RSM exists
whenever $T$ is reached almost surely from \emph{every} state, and this quantifier is
essential. Definition~\ref{def:rsm_for_MDP} imposes the $\varepsilon$-decrease condition at every
$s \in S \setminus T$, so any state that cannot reach $T$ blocks an RSM, and the set of
such states is closed under transitions, so setting $V$ to $0$ there does not help.
\begin{corollary}
\label{cor:existence_almost_sure}
For a finite MDP $M$, fixed policy $\pi$, and target $T \subseteq S$, an
$\varepsilon$-RSM for $(M, \pi)$ w.r.t.\ $T$ exists if and only if
$\mathbb{P}^{\pi}_{s}[\mathrm{Reach}(T)] = 1$ for every $s \in S$.
\end{corollary}
While the forward direction is simply Theorem~\ref{thm:probability_one_if_rsm_exists}, for the converse, the expected hitting time $V(s) = \mathbb{E}_s[\tau]$ is finite because
$S$ is finite, and decomposing on the first step shows it is a $1$-RSM (see sup. mat.~\ref{app:omitted} for details).

Corollary~\ref{cor:existence_almost_sure} underpins the rest of the paper. Certifying
almost-sure reachability is \emph{exactly} the problem of producing an RSM. When $P$ is
known, condition (ii) is directly checkable. When $P$ is unknown, it is an expectation
over an inaccessible kernel, so a candidate can only be evaluated against a learned
model and its verdict carried back to the true MDP. Learning, verifying, and
transferring such certificates are the three questions we take up next.

\section{Learning RSMs under an Unknown Model}
\label{sec:pac_imp}
The objective of this section is to answer a single question. Can one \emph{learn} a
correct RSM certificate, with high probability, from sampling access to an unknown
MDP? We answer in the negative. %
We fix the learning model for almost-sure reachability. A \emph{certificate learning
task} is a tuple $(M,\pi,T)$, where $M=(S,A,P,s_0)$ is an unknown finite MDP with
sampling access, $\pi$ is a fixed policy, and $T\subseteq S$ is a target set.
Certifying $\pi$ with respect to $T$ means establishing
$\mathbb{P}^{\pi}_{s_0}[\mathrm{Reach}(T)]=1$. A learning algorithm $\A$ never sees the
transition kernel and interacts with $M$ only through samples. In each iteration it
either resets the simulator or takes a single step, and outputs its current candidate
RSM certificate $C_i$, inducing a random sequence $\{C_i\}_{i=1}^{\infty}$. This
framework is not specific to reachability. It extends to arbitrary specifications, and
we record the general formulation in sup. mat.~\ref{app:general_learning}.

\begin{definition}[Learning RSM Certificates with Probabilistic Guarantees]
\label{def:learning_rsm_prob_guarantee}
Let $M=(S,A,P,s_0)$ be an unknown finite MDP with sampling access, let $\pi$
be a fixed policy, and let $T\subseteq S$ be a target set. A learning algorithm
$\A$ is said to \emph{learn an RSM certificate with probabilistic guarantees} if
there exists a sample-complexity function $h$ such that, for every $\delta>0$,
taking
\(
N=h\!\left(|S|,|A|,\tfrac{1}{\delta}\right),
\)
we have, for all $n>N$,
$\Pr\bigl[\,C_n \text{ certifies } (M,\pi) \text{ w.r.t.\ } T\,\bigr]\;\geq\;1-\delta.
$
\end{definition}

We claim that no such algorithm can exist, and the reason is already visible in a
three-state MDP as we see next. 
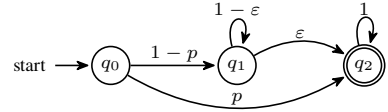
\begin{figure}[t]
\centering
\begin{tikzpicture}[->,>={Stealth[round]},shorten >=1pt,auto,semithick,
                    every state/.style={minimum size=0.5cm,inner sep=1pt,
                                        font=\scriptsize},
                    every loop/.style={min distance=3.5mm,looseness=9},
                    initial text={\scriptsize start}]
  \node[state,initial]   (q0)                                       {$q_0$};
  \node[state]           (q1) [right=0.3cm and 1.15cm of q0]  {$q_1$};
  \node[state,accepting] (q2) [ right=0.3cm and 1.15cm of q1]  {$q_2$};
  \path
    (q0) edge              node[sloped,above,inner sep=1pt] {\scriptsize$1-p$}           (q1)
    (q0) edge[bend right]              node[sloped,below,inner sep=1pt, above] {\scriptsize$p$}             (q2)
    (q1) edge [loop above] node[inner sep=1pt]              {\scriptsize$1-\varepsilon$} (q1)
    (q1) edge [bend left]             node[right,inner sep=1pt, above]        {\scriptsize$\varepsilon$}   (q2)
    (q2) edge [loop above] node[inner sep=1pt]              {\scriptsize$1$}             (q2);
\end{tikzpicture}
\caption{MDP for Lemma~\ref{lem:sample_complexity}}
\label{fig:motivating-mdp}
\end{figure}

\begin{lemma}
\label{lem:sample_complexity}
There exists an unknown finite MDP $M$, a fixed policy $\pi$, and a
reachability specification $\varphi$ such that any learning algorithm $\A$ that
learns RSM certificates with probabilistic guarantees in the sense of
Definition~\ref{def:learning_rsm_prob_guarantee} must have sample complexity depending on
the transition probabilities of $M$. More precisely, for the MDP in Fig.~\ref{fig:motivating-mdp}, %
any such algorithm must use
\[
N
\;\geq\;
1+
\frac{
\log\!\left(\frac{\delta}{1-p}\right)
}{
\log(1-\varepsilon)
}
\]
samples in order to output a correct RSM certificate with probability at least
$1-\delta$.
\end{lemma}

\begin{proof}
At a high level, a correct RSM certificate cannot be produced until
the learner has \emph{witnessed} the sole progress transition, so it suffices to
lower-bound the number of samples needed to witness it. Consider the MDP in Fig.~\ref{fig:motivating-mdp} together with the fixed policy
$\pi$. %
Since $\mathbb{P}^{\pi}_{q_0}[\mathrm{Reach}(q_2)]=1$ whenever
$\varepsilon>0$, an RSM certificate exists, and the task is well posed.
The key observation is that, from samples alone, a correct RSM certificate can be
output only if the transition $q_1\to q_2$ has been observed. Indeed, until that
transition is seen, the observed sub-MDP is indistinguishable from one in which
$q_1$ only self-loops and $q_2$ is unreachable, and in the latter no RSM exists.
Hence the probability of success is at most the probability of the good event
$G_N$ that the transition $q_1\to q_2$ is observed within the first $N$ sampled
transitions. Formally, $\Pr[\text{success}]\le\mathbb{P}[G_N]=1-\mathbb{P}[G_N^c]$.

To lower-bound $\mathbb{P}[G_N^c]$, we exhibit one particular way in which the good
event fails. The process first moves from $q_0$ to $q_1$ (probability $1-p$), and
then takes the self-loop at $q_1$ for each of the remaining $N-1$ transitions
(probability $(1-\varepsilon)^{N-1}$). Therefore,
$\mathbb{P}[G_N^c]\ge(1-p)\,(1-\varepsilon)^{N-1}$.
Consequently, $\mathbb{P}[G_N]\le 1-(1-p)(1-\varepsilon)^{N-1}$. For the success
probability to reach $1-\delta$, we must therefore have
$1-(1-p)\,(1-\varepsilon)^{N-1}\ge 1-\delta$, i.e.,
$(1-p)\,(1-\varepsilon)^{N-1}\le\delta$.
Taking logarithms and dividing by $\log(1-\varepsilon)<0$ (which reverses the
inequality) yields
$(N-1)\,\log(1-\varepsilon)\le\log\!\left(\tfrac{\delta}{1-p}\right)$, i.e.,
$N\ge 1+\log\!\left(\tfrac{\delta}{1-p}\right)\big/\log(1-\varepsilon)$,
as claimed. Note that the bound is informative (it forces $N\ge 2$) exactly when
$1-\varepsilon\ge\frac{\delta}{1-p}$, i.e., in the rare-transition regime where
$\varepsilon$ is small relative to $\frac{\delta}{1-p}$. For $p=0$ the same
argument gives $N\ge\frac{\log(\delta)}{\log(1-\varepsilon)}$.
\end{proof}

Our impossibility result now follows.

\begin{theorem}
\label{thm:pac_imp}
There is no learning algorithm $\A$ that learns RSM certificates for almost-sure
reachability objectives with probabilistic guarantees in the sense of
Definition~\ref{def:learning_rsm_prob_guarantee} using a sample-complexity function
depending only on $|S|$, $|A|$, and $\frac{1}{\delta}$. In particular, for
general unknown finite MDPs, a correct RSM certificate cannot be learned with high
probability efficiently from sampling access alone.
\end{theorem}

\begin{proof}
We argue by contradiction. Suppose such an algorithm $\A$ exists. Then, by
Definition~\ref{def:learning_rsm_prob_guarantee}, there is a sample-complexity function
$h(|S|,|A|,\tfrac{1}{\delta})$ such that, after more than $h$ samples, $\A$
outputs a correct RSM certificate with probability at least $1-\delta$.

However, Lemma~\ref{lem:sample_complexity} exhibits an unknown finite MDP for which
any algorithm that outputs a correct RSM certificate with probability at least
$1-\delta$ must use
$N\ge 1+\log\!\left(\tfrac{\delta}{1-p}\right)\big/\log(1-\varepsilon)$
samples. This lower bound depends on the transition probability $\varepsilon$,
which is unknown and can be chosen arbitrarily small. As $\varepsilon\to 0$, the
bound diverges. Therefore the required number of samples cannot be bounded by any
function of $|S|$, $|A|$, and $\frac{1}{\delta}$ alone, contradicting the assumed
existence of $h$. Hence, in general unknown finite MDPs, it is not possible to
learn correct RSM certificates for almost-sure reachability with high probability
efficiently from sampling access alone.
\end{proof}

Checking a candidate certificate is no easier than producing one. %
To make this precise, we introduce the notion of a sampling-based verifier.

\begin{definition}
\label{def:rsm_verification}
Let $M=(S,A,P,s_0)$ be an unknown finite MDP with sampling access, let $\pi$
be a fixed policy, and let $T\subseteq S$ be a target set. A \emph{sampling-based
verifier} is an algorithm $\mathsf{Ver}$ that, given a candidate function
$V:S\to\mathbb{R}_{\geq 0}$ and sampling access to $M$, decides whether $V$ is
an RSM certificate for $(M,\pi)$ with respect to $T$. We say that $\mathsf{Ver}$
\emph{verifies RSM certificates with probabilistic guarantees} if there exists a
sample-complexity function $h$ such that, for every $\delta>0$, $\mathsf{Ver}$
decides the correctness of $V$ with probability at least $1-\delta$ using at most
\(
h\!\left(|S|,|A|,\tfrac{1}{\delta}\right)
\)
samples.
\end{definition}

\begin{corollary}
\label{cor:verification_impossibility}
For general unknown finite MDPs, there is no sampling-based verifier
$\mathsf{Ver}$ in the sense of Definition~\ref{def:rsm_verification} whose sample
complexity is bounded only by $|S|$, $|A|$, and $\frac{1}{\delta}$.
\end{corollary}

\begin{proof}
By Lemma~\ref{lem:sample_complexity}, even \emph{observing} the progress transition
requires a number of samples that grows without bound as $\varepsilon\to 0$. Since
$\varepsilon$ is unknown, no verifier can distinguish a valid candidate from an
invalid one within a budget depending only on $|S|$, $|A|$, and $\frac{1}{\delta}$,
contradicting existence of such an $h$.
\end{proof}

In summary, the barrier is intrinsic to the unknown model. A single rare transition
can hide for arbitrarily long, and no size-only budget can account for it. 

\section{Learning RSMs under a Known $p_{\min}$}
\label{sec:exp_hit}
\newcommand{\totalerr}{\eta}

The impossibility of Section~\ref{sec:pac_imp} is driven entirely by transitions of
vanishing probability, and assuming a known lower bound $p_{\min}$ on nonzero transition
probabilities suffices to recover an explicit, transferable, and learnable RSM
certificate. This restriction is standard in the literature on learning and statistical
model checking of unknown MDPs~\cite{ashok2019pac}, and it is mild in that it constrains only transitions that are already present. Any nonzero probability must be bounded away from zero by a known amount, while zero-probability transitions are left untouched.

The argument proceeds in four steps. We first adopt the $p_{\min}$ restriction and
show that it makes progress toward the target observable in finitely many samples
(Section~\ref{subsec:p_min-sampling}). We then exhibit a concrete certificate, the
expected hitting-time function, prove it is a valid RSM on every finite MDP
(Theorem~\ref{thm:valid_rsm_proof}), and bound it under $p_{\min}$
(Theorem~\ref{thm:finite_hitting_time}). Next we establish certificate
\emph{transferability}, conditions under which an RSM computed on a learned model
remains valid on the true one (Theorems~\ref{thm:general_certificate_transferability}
and~\ref{thm:transferability_under_pmin}). Finally, we assemble these pieces into an
end-to-end learning guarantee (Theorem~\ref{thm:learn_transferable_eht_rsm}).

\subsection{The $p_{\min}$ Restriction and Observing Progress}
\label{subsec:p_min-sampling}
We define our restriction next.

\begin{definition}[$p_{\min}$]
\label{def:p_min}
Given an MDP $M$, we define the quantity $p_{\min}$ as a known lower bound on every nonzero transition probability of $M$, $p_{\min}\;\le\;\min_{\substack{s\in S,\ a\in \mathrm{Av}(s)\\ t\in \mathrm{Post}(s,a)}} P(s,a,t)$ 
where $\mathrm{Av}(s)\subseteq A$ denotes the set of actions available in $s$, and
$\mathrm{Post}(s,a)=\{t\in S\mid P(s,a,t)>0\}$ denotes the set of possible successors under
action $a$.
\end{definition}

\noindent
Two auxiliary notions make the sampling argument precise, the distance to the target
and the transitions that decrease it. %

\begin{definition}
\label{def:dist_to_target}
Let $M = (S, A, P)$ be a finite MDP, let $\pi$ be a fixed policy, and let
$T \subseteq S$ be a target set.
The \emph{distance function} $\mathrm{Dist} : S \to \mathbb{N}_0 \cup \{\infty\}$ is
defined by
\[
\mathrm{Dist}(s) =
\begin{cases}
0, & \text{if } s \in T, \\[6pt]
1 + \min\limits_{t \in \mathrm{Post}(s, \pi(s))} \mathrm{Dist}(t), & \text{if } s \notin T,
\end{cases}
\]
where $\mathrm{Post}(s, a) = \{\, t \in S \mid P(t \mid s, a) > 0 \,\}$.
\end{definition}

\noindent
Intuitively, $\mathrm{Dist}(s)$ is the minimum number of transitions needed to reach
$T$ from $s$ under $\pi$, with $\mathrm{Dist}(s) = \infty$ when no such path exists.
The recursion has a unique solution in $\mathbb{N}_0 \cup \{\infty\}$, namely
breadth-first distance.

Two conditions recur as hypotheses below, and for a finite MDP they are equivalent.
$ \mathbb{P}^{\pi}_{s}[\mathrm{Reach}(T)] = 1 \ \text{ for every } s \in S
\qquad \Longleftrightarrow \qquad
\mathrm{Dist}(s) < \infty \ \text{ for every } s \in S$. The forward direction
is immediate. $\mathrm{Dist}(s) = \infty$ means no path from $s$
reaches $T$, so $\mathbb{P}^{\pi}_{s}[\mathrm{Reach}(T)] = 0$. For the converse, %
let $q > 0$ be the smallest nonzero transition probability of $M$. Following a
shortest path reaches $T$ within $n - 1$ steps with probability at least
$q^{\,n-1}$ from every state, so the probability of not having reached $T$ after
$k$ blocks of $n - 1$ steps is at most $(1 - q^{\,n-1})^k \to 0$. Note that quantifying over \emph{every} state, rather than over $s_0$ alone, is necessary. The set $\{s : \mathrm{Dist}(s) < \infty\}$ is not in general closed under transitions, so almost-sure reachability from $s_0$ does not propagate to
the states $s_0$ cannot reach (and
Definition~\ref{def:rsm_for_MDP} constrains those states too).

\begin{definition}
\label{def:progress_making}
Let $s \in S \setminus T$, let $a = \pi(s)$, and let $t \in S$. A transition from
$s$ to $t$ is a \emph{progress-making transition} if
$P(t \mid s, a) > 0$
and 
$\mathrm{Dist}(t) = \mathrm{Dist}(s) - 1$.
\end{definition}

\noindent
A progress-making transition strictly reduces the distance to the target. The key
underpinning observation is that, once every realizable transition carries
probability at least $p_{\min}$, such a transition can no longer carry an
arbitrarily small probability, so progress toward $T$ becomes observable with high
probability in finitely many samples.

\begin{lemma}
\label{lem:pmin_progress_path_sampling}
Let $M=(S,A,P,s_0)$ be an unknown finite MDP, let $\pi$ be a fixed policy, and let
$T\subseteq S$ be a target set with $\mathbb{P}_{s}^{\pi}[\mathrm{Reach}(T)]=1$ for
every $s\in S$.
Suppose a lower bound $p_{\min}>0$ on every nonzero transition probability is
known. Then, with probability at least $1-\delta$, a path to $T$ can be observed
using a number of samples bounded in terms of $|S|$, $p_{\min}$, and $\delta$.
Here, observing a path to $T$ means observing a finite sample trajectory
$s_0,s_1,\ldots,s_k$ under $\pi$ with $s_k\in T$.
\end{lemma}

\begin{proof}[Proof sketch]
The strategy is to bound the length of a shortest progress path, lower-bound the
probability of following it, and then amplify by independent episodes. Under $\pi$,
every reachable state has a shortest path to $T$ of length
$\operatorname{Dist}(s)\le n-1$, and each step of such a path is a progress
transition, which by the $p_{\min}$ assumption has probability at least $p_{\min}$.
Hence from any reachable state $T$ is reached within
$D=\max_{s}\operatorname{Dist}(s)$ steps with probability at least
$p:=(p_{\min})^{D}$. Running independent episodes of length $D$, the failure
probability decays as $(1-p)^m$, so $m\ge\log(\delta)/\log(1-p)$ episodes (that is,
$N=mD\ge(n-1)\cdot\log(\delta)/\log\!\big(1-(p_{\min})^{n-1}\big)$ samples) drive it
below $\delta$. The full argument is given in sup. mat.~\ref{app:omitted}.
\end{proof}

Note that the role of
$p_{\min}$ is to simply remove the rare-transition
barrier of Lemma~\ref{lem:sample_complexity}, so that progress toward the target is
observed with probability $1-\delta$ within finitely many samples. Observing
progress, however, is not the same as learning or verifying a certificate. Both
additionally require the expected-decrease inequality
$V(s)\ge\mathbb{E}_{t\sim P(s,\pi(s))}[V(t)]+\varepsilon$ to hold at every
non-target $s$, and since $P$ is unknown, these expectations must themselves be
estimated. The remainder of this section closes exactly this gap. We construct such
a $V$, bound it, and show that it transfers from the learned model to the true one.

\subsection{The Expected Hitting-Time Certificate}
\label{subsec:eht}
We now exhibit the concrete certificate.

\begin{definition}
\label{def:exp_hit_time}
Let $M = (S, A, P)$ be an MDP, let $\pi$ be a policy, and let $T \subseteq S$ be a
target set. The \emph{hitting time} of $T$ is defined as
$H_T := \min \{ t \ge 0 \mid X_t \in T \},$
where $X_t$ denotes the state at time step $t$. The \emph{expected hitting time}
from a state $s \in S$ under $\pi$ is then defined as
\[
\mathbb{E}_s^\pi[H_T] =
\begin{cases}
0, & \text{if } s \in T, \\[6pt]
1 + \sum_{y \in S} P(y \mid s, \pi(s))\, \mathbb{E}_y^\pi[H_T], & \text{if } s \notin T.
\end{cases}
\]
\end{definition}

\noindent
We adopt the recursive form deliberately. Each application removes exactly one
step, and this ``$+1$'' supplies the strict expected decrease an RSM requires. The
expected hitting time is therefore a natural RSM candidate, as we show next.%

\begin{theorem}
\label{thm:valid_rsm_proof}
    Let $M = (S, A, P, s_0)$ be a finite MDP, let $\pi$ be a fixed policy, and let
    $T \subseteq S$ be a target set that is almost surely reachable from every
    state, i.e.\ $\mathbb{P}^{\pi}_{s}[\mathrm{Reach}(T)] = 1$ for every $s \in S$.
    Define $V : S \to \mathbb{R}_{\ge 0}$ by
  $      V(s) = \mathbb{E}_s^\pi[H_T]$.
    Then $V$ is an $\varepsilon$-RSM with $\varepsilon = 1$.
\end{theorem}

\begin{proof}[Proof sketch]
Both RSM conditions follow directly from the recursive form of
Definition~\ref{def:exp_hit_time}. Non-negativity is immediate, since
$V(s)=\mathbb{E}_s^\pi[H_T]$ is the expectation of a non-negative random variable.
For expected decrease, the recursion gives
$V(s)=1+\mathbb{E}_{t\sim P(s,\pi(s))}[V(t)]$ at every non-target $s$, which is
exactly the $\varepsilon$-decrease condition with $\varepsilon=1$. %
\end{proof}

We now bound the expected hitting time under the $p_{\min}$ restriction, using the
distance function and progress-making transitions from
Section~\ref{subsec:p_min-sampling}. The bound is deliberately worst-case. It requires
no knowledge of the model beyond $p_{\min}$, at the cost of being loose on any
particular instance.

\begin{theorem}
\label{thm:finite_hitting_time}
Let $M = (S, A, P, s_0)$ be a finite MDP, let $\pi$ be a fixed policy, and let
$T \subseteq S$ be a target set that is reachable almost surely under $\pi$ from
every state. Let $p_{\min} > 0$ be the minimum nonzero transition probability, and
let $D = \max_{s \in S} \mathrm{Dist}(s)$, which is finite by
Definition~\ref{def:dist_to_target}. Then, for every state $s \in S$, the expected
hitting time satisfies
$\mathbb{E}_s^\pi[H_T] \le \frac{D}{(p_{\min})^D}.$
\end{theorem}

\begin{proof}[Proof sketch]
The strategy is to reach $T$ within $D$ steps with probability at least
$(p_{\min})^D$, and then turn this per-block success probability into a geometric
bound on the expected hitting time. Fix a state $s$ with $\mathrm{Dist}(s) < \infty$.
Following progress-making transitions, each of probability at least $p_{\min}$
(Definition~\ref{def:progress_making}), reaches $T$ within
$D = \max_x \mathrm{Dist}(x)$ steps with probability at least $p := (p_{\min})^D$.
Dividing time into blocks of length $D$, the target is missed for $k$ consecutive
blocks with probability at most $(1-p)^k$, so
$\mathbb{E}_s^\pi[H_T] = \sum_{t \ge 0} \mathbb{P}_s^\pi[H_T > t]
\le D \sum_{k \ge 0} (1-p)^k = D/p = D/(p_{\min})^D$.
See sup. mat.~\ref{app:omitted} for details.
\end{proof}

\noindent
The inequality cannot be strengthened to a strict one. Take
$S = \{s_0, t\}$ with $T = \{t\}$, a single action, and
$P(t \mid s_0, \pi(s_0)) = P(s_0 \mid s_0, \pi(s_0)) = \tfrac12$. Then
$p_{\min} = \tfrac12$, $\mathrm{Dist}(s_0) = 1$ so $D = 1$, and
$\mathbb{E}_{s_0}^\pi[H_T] = 2 = D/(p_{\min})^D$. (The bound is also attained
degenerately when $T = S$, where $D = 0$ and both sides are $0$.)

\subsection{Transferability}
\label{subsec:transferability}
The central
question of this section is whether an RSM certificate computed on a \emph{learned model}
$M'$ remains valid on the true model $M$? A learned model $M' = (S, A, P', s_0)$
is an MDP with the same states, actions, and initial state as $M$, but with
transition probabilities $P'$ estimated from samples. %
Then we see that validity does not transfer automatically. Since $V$ depends on the state only, its
values are identical on both models. The difficulty lies in the \emph{expectation}, which the two
models take with respect to different transition probabilities. The expected-decrease
condition could in principle fail on the true model even though the $V$-values agree.
This threat, however, is controllable. We prove that a sufficiently close
approximation preserves the RSM property, so the certificate transfers. We first
quantify the approximation error.

\begin{definition}
\label{def:total_prob_error}
Let $M = (S, A, P, s_0)$ be the original MDP and let $M' = (S, A, P', s_0)$ be the learned
MDP. The \emph{total probability error} $\totalerr$ is defined as
\[
\totalerr = \sum_{x \in S} \sum_{a \in A} \sum_{y \in S}
\left| P(y \mid x, a) - P'(y \mid x, a) \right|.
\]
We say that $M'$ is an \emph{$\eta$-approximation} of $M$, for $\eta > 0$, if the
total probability error between the two is at most $\eta$.
\end{definition}

\noindent
We now establish certificate transferability in full generality, without any
restriction on the MDP.

\begin{theorem}
\label{thm:general_certificate_transferability}
Let $M = (S, A, P, s_0)$ be a finite MDP, let $M' = (S, A, P', s_0)$ be an
$\eta$-approximation of $M$, and let $\pi$ be a fixed policy. Let
$T \subseteq S$ be a target set. Suppose $V : S \to \mathbb{R}_{\ge 0}$ is an
$\varepsilon'$-RSM for $(M', \pi)$ with respect to $T$.
Assume $V$ is bounded on $S$, and let $B := \max_{s \in S} V(s) < \infty$. If
$\eta < \frac{\varepsilon'}{B}$,
then $V$ is an $\varepsilon$-RSM for $M$ with
$\varepsilon = \varepsilon' - B\eta > 0$.
\end{theorem}

\begin{proof}
The key idea is to bound the discrepancy between the one-step expectations of $V$
on the two models by $B\eta$, and then absorb it into the slack $\varepsilon'$ of
the learned-model certificate. Fix $x \in S$ and let $a = \pi(x)$. By definition
of expectation,
$\mathbb{E}_M[V(X_{t+1}) \mid x, a] = \sum_y P(y \mid x, a)\, V(y)$ and
$\mathbb{E}_{M'}[V(X_{t+1}) \mid x, a] = \sum_y P'(y \mid x, a)\, V(y)$.
Subtracting and applying the triangle inequality,
$\mathbb{E}_M[V(X_{t+1}) \mid x, a] - \mathbb{E}_{M'}[V(X_{t+1}) \mid x, a]
\le \sum_y |P(y \mid x, a) - P'(y \mid x, a)| \cdot |V(y)|$.
Since $|V(y)| \le B$ for all $y \in S$, and since the $\eta$-approximation
assumption gives $\sum_y |P(y \mid x, a) - P'(y \mid x, a)| \le \eta$, we obtain
$\mathbb{E}_M[V(X_{t+1}) \mid x, a]
\le \mathbb{E}_{M'}[V(X_{t+1}) \mid x, a] + B\eta$.
Because $V$ is an $\varepsilon'$-RSM for $(M', \pi)$, we have
$\mathbb{E}_{M'}[V(X_{t+1}) \mid x, a] \le V(x) - \varepsilon'$. Substituting,
$\mathbb{E}_M[V(X_{t+1}) \mid x, a] \le V(x) - \varepsilon' + B\eta
= V(x) - (\varepsilon' - B\eta)$.
Setting $\varepsilon := \varepsilon' - B\eta$, this is exactly the expected
decrease condition on $M$, and $\varepsilon > 0$ precisely when
$\eta < \varepsilon' / B$. Hence $V$ is an $\varepsilon$-RSM for $M$.
\end{proof}

\noindent
The condition of
Theorem~\ref{thm:general_certificate_transferability} is not immediately checkable.
It couples three entangled quantities $(\eta, B, \varepsilon')$. Deriving $B$
requires an approximate model, which itself carries error $\eta$, while
$\varepsilon'$ depends on the computed certificate. The condition becomes checkable a
priori once two of the three are fixed in advance, which is exactly what the
$p_{\min}$ restriction and the expected hitting-time certificate provide.
Theorem~\ref{thm:finite_hitting_time} bounds $B$, and
Theorem~\ref{thm:valid_rsm_proof} gives $\varepsilon' = 1$. One subtlety remains.
Here $B$ is a supremum of expected hitting times \emph{of the learned model}
$M'$, whereas $p_{\min}$ bounds the nonzero transition probabilities of the
\emph{true} model $M$, so Theorem~\ref{thm:finite_hitting_time} does not apply to
$M'$ with $p_{\min}$ directly. It applies with $p_{\min}/2$. Once
$\eta \le p_{\min}/2$, every transition of probability at least $p_{\min}$
in $M$ retains probability at least $p_{\min} - \eta \ge p_{\min}/2$ in $M'$, so
the progress paths of Definition~\ref{def:progress_making} survive into $M'$ with
edge weight $p_{\min}/2$. With $B$ and $\varepsilon'$ pinned down, one refines the
approximation until $\eta$ falls below the transfer threshold. We now specialize
Theorem~\ref{thm:general_certificate_transferability} to the expected hitting-time
certificate, for which $\varepsilon' = 1$.

\begin{theorem}
\label{thm:transferability_under_pmin}
Let $M' = (S, A, P', s_0)$ be an $\eta$-approximation of a finite MDP
$M = (S, A, P, s_0)$, let $\pi$ be a fixed policy, and let $T \subseteq S$ be
reachable almost surely under $\pi$ from every state. Let
$V(s) = \mathbb{E}^{\pi, M'}_{s}[H_T]$ be
the expected hitting-time certificate of $M'$ (a $1$-RSM for $(M', \pi)$ by
Theorem~\ref{thm:valid_rsm_proof}), and let $B := \max_{s \in S} V(s)$. If
$\eta < 1/B$, then $V$ is an $\varepsilon$-RSM for $M$ with
$\varepsilon = 1 - B\eta$. Moreover, if a lower bound $p_{\min} > 0$ is known and
$D = \max_{x:\, \mathrm{Dist}(x) < \infty} \mathrm{Dist}(x) \ge 1$, then
$\eta < (p_{\min}/2)^D/D$ suffices, and it gives
$B \le D/(p_{\min}/2)^D$.
\end{theorem}

\begin{proof}[Proof sketch]
This is Theorem~\ref{thm:general_certificate_transferability} specialized to
$\varepsilon' = 1$. The expected hitting-time function is a $1$-RSM on $M'$
(Theorem~\ref{thm:valid_rsm_proof}) with $|V| \le B$, so the one-step expectations
of $V$ under $M$ and $M'$ differ by at most $B\eta$, and the unit slack survives as
$\varepsilon = 1 - B\eta > 0$ whenever $\eta < 1/B$. The bound on $B$ then follows
from Theorem~\ref{thm:finite_hitting_time}. %
\end{proof}

\subsection{The Learning Guarantee}
\label{subsec:learning_guarantee}
We conclude by assembling these pieces into an end-to-end learning guarantee. We
begin from the standard result that an $\eta$-close approximation can be learned.

\begin{lemma}[Learning an MDP approximation \cite{brafman2001rmax,azar2013minimax}]
\label{lem:learn_close_mdp}
Let $\M = (S, A, P, s_0)$ be an unknown finite MDP with sampling access. For every
accuracy parameter $\alpha > 0$ and confidence parameter $\delta > 0$, there
exists a model-learning procedure that returns a learned MDP
$\M' = (S, A, P', s_0)$ with
$\eta \le \alpha$
with probability at least $1 - \delta$, using a number of samples polynomial in
$|S|$, $|A|$, $\tfrac{1}{\alpha}$, and $\log\tfrac{1}{\delta}$.
\end{lemma}

\begin{theorem}
\label{thm:learn_transferable_eht_rsm}
Let $M$ be an unknown finite MDP, $\pi$ a fixed policy, and $T$ be a
target set reached almost surely under $\pi$ from every state. Suppose a lower
bound
$p_{\min} > 0$ on every nonzero transition probability is known, and let
$D = \max\{\mathrm{Dist}(s) : \mathrm{Dist}(s) < \infty\}$. Then, for every
$\delta > 0$, an expected hitting-time RSM certificate for $(M, \pi)$ w.r.t $T$ can be learned with probability at least $1 - \delta$, using a number of
samples polynomial in $|S|$, $|A|$, $\tfrac{D}{(p_{\min}/2)^{D}}$ and
$\log\tfrac{1}{\delta}$.
\end{theorem}

\begin{proof}
The high-level strategy is to learn a model accurate enough that the transfer
condition of Theorem~\ref{thm:transferability_under_pmin} is met, compute the
expected hitting-time certificate on that model, and invoke transferability. Let
$D = \max\{\mathrm{Dist}(s) : s \in S,\ \mathrm{Dist}(s) < \infty\}$ and
$\alpha = (p_{\min}/2)^D/(2D)$.
By Lemma~\ref{lem:learn_close_mdp}, with probability at least $1 - \delta$ we can
learn an MDP approximation $M' = (S, A, P', s_0)$ with $\eta < \alpha$. On $M'$,
define the expected hitting-time function $V(s) = \mathbb{E}^{\pi, M'}_s[H_T]$.
By Theorem~\ref{thm:valid_rsm_proof}, $V$ is a $1$-RSM for $(M', \pi)$ with respect to
$T$, and by the expected hitting-time bound under the $p_{\min}$ assumption
(Theorem~\ref{thm:finite_hitting_time}, applied to $M'$ with $p_{\min}/2$ as in
Theorem~\ref{thm:transferability_under_pmin}), $B := \max_{s \in S} V(s) \le
\tfrac{D}{(p_{\min}/2)^D}$. Note that $\alpha \le p_{\min}/2$ for $D \ge 1$, so
the hypothesis $\eta < \alpha$ already supplies the condition that bound needs.
Hence $\eta < \alpha < (p_{\min}/2)^D/D \le 1/B$,
so by Theorem~\ref{thm:transferability_under_pmin} the certificate $V$ transfers from
$M'$ to $M$. Consequently, with probability at least $1 - \delta$, $V$ is a valid
RSM certificate for $(\M, \pi)$ with respect to $T$.

It remains to account for the computational cost. The complexity of learning $M'$ is
established by Lemma~\ref{lem:learn_close_mdp}. Given $M'$, the expected hitting-time
function $V$ is computed by solving the linear system
$V(s) = 1 + \sum_{t \in S} P'(t \mid s, \pi(s))\, V(t)$ for $s \in S \setminus T$,
with boundary condition $V(s) = 0$ for all $s \in T$. Writing $n = |S|$, this
system has at most $n$ variables and is solved by Gaussian elimination in
$O(n^3)$ time, a cost dominated by the sample complexity of
Lemma~\ref{lem:learn_close_mdp}. Therefore, the overall complexity of learning a
correct expected-hitting-time RSM is governed by the complexity of learning an
$\eta$-approximation of the MDP at accuracy $\alpha$, and is in particular
polynomial in $|S|$, $|A|$, $\tfrac{1}{\alpha} = \tfrac{2D}{(p_{\min}/2)^{D}}$ and
$\log\tfrac{1}{\delta}$, which proves the theorem.
\end{proof}

Theorem~\ref{thm:learn_transferable_eht_rsm} delivers a certificate with
probabilistic guarantees, but it is deliberately \emph{not} stated as an
efficiency claim. The
accuracy the model learner must be run at is $\alpha = (p_{\min}/2)^{D}/(2D)$, so
the sample complexity is polynomial in
$1/\alpha = 2D\,(2/p_{\min})^{D}$, which is exponential in $D$. Moreover $D$ is not available to the learner (it is
determined by the support of the unknown kernel $P$), so the only a priori
substitute is the worst-case bound $D \le |S| - 1$. Whether the exponential
dependence on $D$ is necessary, or is an artifact of the worst-case bound of
Theorem~\ref{thm:finite_hitting_time}, we leave open. Our experiments %
indicate that this bound is loose in practice.

\section{Empirical Evaluation of Transferability}
\label{sec:experiments}

The objective of this section is to put our transfer guarantee to an empirical test.
Recall that Theorem~\ref{thm:learn_transferable_eht_rsm} is only \emph{sufficient}. An
EHT RSM $V$ computed on a learned model $M'$ is 
valid on the true MDP $M$ when $\varepsilon' = 1 - B\eta > 0$, where
$B = \max_{s} V(s)$ and $\eta$ is the approximation error. Since the guarantee is
one-directional, two questions  arise. First, how demanding is the condition
$\varepsilon' > 0$ in practice? Second, does the certificate ever transfer even when
the condition fails?

\paragraph{Setup.}
For each benchmark we compute $V$ on the learned model $M'$ (a $1$-RSM for
$(M',\pi)$ by Theorem~\ref{thm:valid_rsm_proof}), perturb the kernel to obtain the
true MDP $M$, and record two quantities. The first is the sufficient condition
$\varepsilon' = 1 - B\eta$. The second is the \emph{empirical drift}
$\mathrm{drift}_{\min} = \min_{s}\big(V(s) - \mathbb{E}_M[V(\text{next})]\big)$. The
first is what our theory can certify, namely $\varepsilon' > 0$. The second is the
ground truth, since $V$ is a valid certificate on $M$ exactly when
$\mathrm{drift}_{\min} > 0$.

\paragraph{Benchmarks.}
We evaluate on seven MDPs spanning a range of progress dynamics. 
These are
three cyclic MDPs (forward probability $0.01$, a single $0.01$ bottleneck, and $0.1$),
a DAG-like MDP, a gridworld, a RiverSwim-style chain, and a large sparse $100$-state
MDP. The suite  mixes benign structures with ones containing rare
progress transitions, since the rare transitions are precisely what inflate the
hitting-time bound $B$. All benchmark MDPs, their exact kernels, and per-state certificate and drift values are given in sup. mat.~\ref{app:benchmarks}, so every value in Table~\ref{tab:experiments} is reproducible without code.

\begin{table}[t]
\centering
\small
\setlength{\tabcolsep}{5pt}
\begin{tabular}{@{}lrrrr@{}}
\toprule
Benchmark & $\eta$ & $B$ & $\varepsilon'$ & drift$_{\min}$ \\
\midrule
Cyclic ($p{=}0.01$)   & $0.0167$ & $200$      & $-2.333$    & $0.500$ \\
Cyclic (bottleneck)   & $0.020$  & $82.5$     & $-0.650$    & $0.783$ \\
Cyclic ($p{=}0.1$)    & $0.020$  & $20$       & $\phantom{-}0.600$ & $0.933$ \\
DAG-like              & $0.010$  & $3.132$    & $\phantom{-}0.969$ & $0.999$ \\
Gridworld             & $0.010$  & $5.001$    & $\phantom{-}0.950$ & $0.998$ \\
Chain (RiverSwim)     & $0.010$  & $15.695$   & $\phantom{-}0.843$ & $0.997$ \\
Large $100$-state     & $0.010$  & $10666.7$  & $-105.667$  & $0.958$ \\
\bottomrule
\end{tabular}
\caption{Transferability across benchmarks, with $\varepsilon' = 1 - B\eta$ and
drift$_{\min} = \min_{s}\big(V(s) - \mathbb{E}_M[V(\text{next})]\big)$. The
certificate is empirically valid on $M$ (drift$_{\min} > 0$) in \emph{every}
case, yet the sufficient condition $\varepsilon' > 0$ fails wherever a rare
transition inflates $B$.}
\label{tab:experiments}
\end{table}

\paragraph{Observations.}
Two findings answer our questions. First, the condition is easy to meet exactly when
the MDP is progress-making. For the DAG-like, gridworld, chain, and $p{=}0.1$ cyclic
models, forward probabilities are not vanishingly small, so $B$ stays small and
$\varepsilon'$ is comfortably positive, indeed close to $1$. Second, a few rare
transitions break it. For the all-$0.01$ cyclic MDP a mere $\eta = 0.0167$ already
forces $\varepsilon' = -2.333$ (through $B = 200$), and the $100$-state model is the
extreme case, where a single rare transition inflates $B$ to $10666.7$ and
$\varepsilon' = -105.667$.

The decisive column, however, is drift$_{\min}$. It is strictly positive in
\emph{every} benchmark (never below $0.5$), including all three where
$\varepsilon' < 0$. In other words, the certificate transfers in practice even where
our theory cannot promise it. This confirms that $1 - B\eta > 0$, while sound, is
conservative. It multiplies the \emph{global} worst-case bound $B$ against $\eta$, so
a single rare state can inflate $B$ without endangering the drift elsewhere. Improving this gap is part of future work. %

\section{Conclusion}
We showed that almost-sure reachability cannot be certified from samples alone in general unknown MDPs (Section~\ref{sec:pac_imp}), yet a known bound $p_{\min}$ recovers a learnable, checkable, and transferable expected-hitting-time certificate (Section~\ref{sec:exp_hit}). As future work, we plan to investigate whether the exponential dependence on $D$ (Section~\ref{sec:experiments}) can be improved, and whether a similar certificate-based approach extends to reach-avoid and full LTL specifications.

\bibliography{refs}

\clearpage
\appendix
\section{Preliminaries}

\subsection{Measure-Theoretic Background}
\label{app:sigma_algebra}
For completeness, we collect here the measure-theoretic notions underlying the
stochastic-process language used in Section~\ref{sec:prelim}. None of this is
new; we follow standard treatments~\cite{williams1991probability}. Readers
familiar with basic probability theory may safely skip this supplementary material.

\paragraph{$\sigma$-algebras.}
Let $\Omega$ be a nonempty set, thought of as the \emph{sample space} of all
possible outcomes of a random experiment. A collection
$\mathcal{F} \subseteq 2^{\Omega}$ of subsets of $\Omega$ is a
\emph{$\sigma$-algebra} (or \emph{$\sigma$-field}) over $\Omega$ if it is closed
under complementation and countable unions and contains the whole space:
\begin{enumerate}
  \item $\Omega \in \mathcal{F}$;
  \item if $A \in \mathcal{F}$, then its complement $A^{c} = \Omega \setminus A
        \in \mathcal{F}$; and
  \item if $A_1, A_2, A_3, \ldots \in \mathcal{F}$, then
        $\bigcup_{i=1}^{\infty} A_i \in \mathcal{F}$.
\end{enumerate}
These three conditions also force $\emptyset \in \mathcal{F}$ (by (1) and (2))
and closure under countable intersections (by (2), (3), and De Morgan's laws).
The elements of $\mathcal{F}$ are called \emph{events}: a $\sigma$-algebra is
thus the collection of subsets of $\Omega$ to which we are able to assign a
probability. The pair $(\Omega, \mathcal{F})$ is called a \emph{measurable
space}.

Two extreme examples are the \emph{trivial} $\sigma$-algebra
$\{\emptyset, \Omega\}$, which records no information, and the \emph{power set}
$2^{\Omega}$, which records everything and is the default choice when $\Omega$ is
countable (as for the finite MDPs of this paper). Intermediate $\sigma$-algebras
model \emph{partial} information, which is precisely what filtrations exploit.

\paragraph{Generated $\sigma$-algebras.}
Given any family $\mathcal{C} \subseteq 2^{\Omega}$ of subsets, the
\emph{$\sigma$-algebra generated by $\mathcal{C}$}, denoted
$\sigma(\mathcal{C})$, is the smallest $\sigma$-algebra containing
$\mathcal{C}$, i.e.\ the intersection of all $\sigma$-algebras that contain
$\mathcal{C}$. This intersection is itself a $\sigma$-algebra and is well defined
because $2^{\Omega}$ is one such $\sigma$-algebra, so the family being
intersected is nonempty. Analogously, for a random variable $X$ we write
$\sigma(X)$ for the smallest $\sigma$-algebra with respect to which $X$ is
measurable; intuitively, $\sigma(X)$ encodes exactly the information revealed by
observing $X$.

\paragraph{Probability measures and probability spaces.}
A \emph{probability measure} on $(\Omega, \mathcal{F})$ is a function
$\mathbb{P} : \mathcal{F} \to [0,1]$ satisfying
\begin{enumerate}
  \item \textbf{Normalization:} $\mathbb{P}(\Omega) = 1$; and
  \item \textbf{Countable additivity:} for any countable collection of pairwise
        disjoint events $A_1, A_2, A_3, \ldots \in \mathcal{F}$,
        \[
          \mathbb{P}\!\left(\bigcup_{i=1}^{\infty} A_i\right)
          = \sum_{i=1}^{\infty} \mathbb{P}(A_i).
        \]
\end{enumerate}
The triple $(\Omega, \mathcal{F}, \mathbb{P})$ is a \emph{probability space}. It
is the fundamental object on which all random quantities in the paper are
defined.

\paragraph{Random variables and expectation.}
A \emph{random variable} is a measurable function $X : \Omega \to \mathbb{R}$,
meaning that $X^{-1}(B) \in \mathcal{F}$ for every Borel set
$B \subseteq \mathbb{R}$; measurability guarantees that events such as
$\{X \le x\} = X^{-1}((-\infty, x])$ lie in $\mathcal{F}$ and hence have a
well-defined probability. We write $\mathbb{E}[X]$ for the expectation
(Lebesgue integral) of $X$ with respect to $\mathbb{P}$, when it exists, and
$\mathbb{E}[X \mid \mathcal{G}]$ for the conditional expectation of $X$ given a
sub-$\sigma$-algebra $\mathcal{G} \subseteq \mathcal{F}$, i.e.\ the (almost
surely unique) $\mathcal{G}$-measurable random variable satisfying
$\mathbb{E}[\,\mathbb{E}[X \mid \mathcal{G}]\,\mathbf{1}_G\,]
 = \mathbb{E}[X\,\mathbf{1}_G]$ for all $G \in \mathcal{G}$. Conditional
expectation is the formal notion of ``best prediction of $X$ given the
information in $\mathcal{G}$,'' and it is the operation appearing in the defining
inequality of a supermartingale (Definition~\ref{def:supermartingales}).

\paragraph{Stochastic processes and filtrations.}
A \emph{(discrete-time) stochastic process} is a sequence $(X_k)_{k \ge 0}$ of
random variables on a common probability space
$(\Omega, \mathcal{F}, \mathbb{P})$. A \emph{filtration} is a nondecreasing
sequence of sub-$\sigma$-algebras
\[
  \mathcal{F}_0 \subseteq \mathcal{F}_1 \subseteq \mathcal{F}_2
  \subseteq \cdots \subseteq \mathcal{F},
\]
where $\mathcal{F}_k$ models the information available up to time $k$; the
nesting expresses that information accumulates and is never forgotten. A process
$(X_k)$ is \emph{adapted} to $(\mathcal{F}_k)$ if each $X_k$ is
$\mathcal{F}_k$-measurable, i.e.\ its value is determined by the information
available at time $k$. The canonical choice is the \emph{natural filtration}
$\mathcal{F}_k = \sigma(X_0, \ldots, X_k)$, generated by the history of the
process itself. For an MDP under a fixed policy, this is exactly the
$\sigma$-algebra $\sigma(s_0, \ldots, s_k)$ generated by the observed trajectory
prefix, which is the filtration used in the main text.

\tikzset{
  mdp/.style={->,>=Stealth,shorten >=1pt,auto,semithick,
              every state/.style={minimum size=7mm,inner sep=1pt,font=\small}},
  plab/.style={font=\footnotesize},
}

\section{Benchmark Details}
\label{app:benchmarks}

This supplementary material supports the empirical evaluation of Section~\ref{sec:experiments}
with the material that does not fit in the main text: for each benchmark we give
the exact transition kernel of the true MDP $M$, drawn as an automaton, together
with the per-state certificate and drift values behind the aggregate results.
The approximation error $\eta$, the bound $B$, the transfer margin
$\varepsilon' = 1 - B\eta$, and the minimum drift are defined in
Section~\ref{sec:experiments}, and their per-benchmark values are collected in
Table~\ref{tab:experiments}; we do not repeat them here. Instead, each table
below reports, per state, the certificate value $V(s)$, its one-step
expectations $\mathbb{E}_M[V(\text{next})]$ and $\mathbb{E}_{M'}[V(\text{next})]$
under the true and nominal kernels, and the corresponding drifts $\mathrm{Drift}(M)$
and $\mathrm{Drift}(M')$. As in the main text, the larger benchmarks are shown on
a representative subset of states. All reported quantities are computed by exact linear-algebra solves, since the expected hitting time is the unique solution of a linear system. The results are therefore deterministic, involve no randomness, and complete in negligible time on a standard laptop.

\subsection{Cyclic MDP, all forward probabilities $0.01$}
\label{app:cyc001}
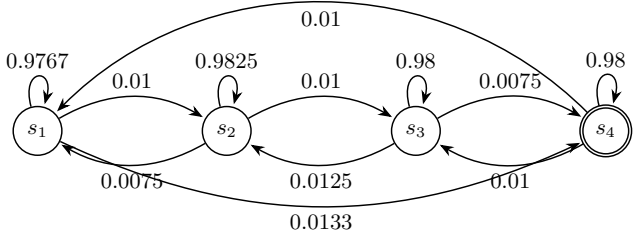
\begin{figure}[t]
\centering
\resizebox{\columnwidth}{!}{%
\begin{tikzpicture}[mdp,node distance=2cm]
\node[state] (s1) {$s_1$};
\node[state,right=of s1] (s2) {$s_2$};
\node[state,right=of s2] (s3) {$s_3$};
\node[state,accepting,right=of s3] (s4) {$s_4$};
\path[plab]
 (s1) edge[loop above] node {$0.9767$} (s1)
 (s2) edge[loop above] node {$0.9825$} (s2)
 (s3) edge[loop above] node {$0.98$} (s3)
 (s4) edge[loop above] node {$0.98$} (s4)
 (s1) edge[bend left] node {$0.01$} (s2)
 (s2) edge[bend left] node {$0.0075$} (s1)
 (s2) edge[bend left] node {$0.01$} (s3)
 (s3) edge[bend left] node {$0.0125$} (s2)
 (s3) edge[bend left] node {$0.0075$} (s4)
 (s4) edge[bend left] node {$0.01$} (s3)
 (s1) edge[bend right=25] node[below] {$0.0133$} (s4)
 (s4) edge[bend right=45] node[below] {$0.01$} (s1);
\end{tikzpicture}}
\caption{Cyclic MDP with all forward probabilities $0.01$ (true kernel $M$).}
\label{fig:cyc001}
\end{figure}

This benchmark (Fig.~\ref{fig:cyc001}) is a four-state cyclic MDP
$s_1,\dots,s_4$ with target $s_4$; every state overwhelmingly self-loops, and
progress is made only through rare forward transitions of order $0.01$. These
rare transitions inflate the expected hitting time, and hence the bound $B$,
which is what drives the transfer margin negative for this instance even though
the true-MDP drift stays positive (per-state values in Table~\ref{tab:cyc001}).

\begin{table}[t]
\centering
\small
\setlength{\tabcolsep}{4pt}
\begin{tabular}{@{}lrrrrr@{}}
\toprule
State & $V(s)$ & $\mathbb{E}_M[V']$ & $\mathbb{E}_{M'}[V']$ & Drift$(M)$ & Drift$(M')$ \\
\midrule
$s_0$ & $150$ & $148.5$   & $149$ & $1.5$   & $1$ \\
$s_1$ & $200$ & $199.125$ & $199$ & $0.875$ & $1$ \\
$s_2$ & $150$ & $149.5$   & $149$ & $0.5$   & $1$ \\
\bottomrule
\end{tabular}
\caption{Per-state certificate and drift values for the cyclic MDP with all
forward probabilities $0.01$.}
\label{tab:cyc001}
\end{table}

\subsection{Cyclic MDP, one bottleneck forward probability $0.01$}
\label{app:cycbottle}
\begin{figure}[t]
\centering
\resizebox{\columnwidth}{!}{%
\begin{tikzpicture}[mdp,node distance=2cm]
\node[state] (s1) {$s_1$};
\node[state,right=of s1] (s2) {$s_2$};
\node[state,right=of s2] (s3) {$s_3$};
\node[state,accepting,right=of s3] (s4) {$s_4$};
\path[plab]
 (s1) edge[loop above] node {$0.9767$} (s1)
 (s2) edge[loop above] node {$0.8033$} (s2)
 (s3) edge[loop above] node {$0.8$} (s3)
 (s4) edge[loop above] node {$0.8$} (s4)
 (s1) edge[bend left] node {$0.01$} (s2)
 (s2) edge[bend left] node {$0.0967$} (s1)
 (s2) edge[bend left] node {$0.1$} (s3)
 (s3) edge[bend left] node {$0.1033$} (s2)
 (s3) edge[bend left] node {$0.0967$} (s4)
 (s4) edge[bend left] node {$0.1$} (s3)
 (s1) edge[bend right=25] node[below] {$0.0133$} (s4)
 (s4) edge[bend right=45] node[below] {$0.1$} (s1);
\end{tikzpicture}}
\caption{Cyclic MDP with one bottleneck forward transition of probability $0.01$
(true kernel $M$).}
\label{fig:cycbottle}
\end{figure}
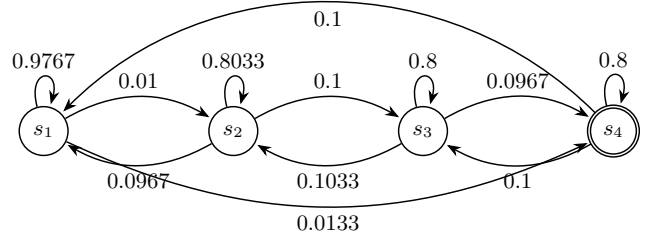

This variant (Fig.~\ref{fig:cycbottle}) keeps the four-state cyclic structure but
makes only a \emph{single} forward transition rare, with the remaining forward
probabilities raised to $0.1$. The lone bottleneck alone suffices to inflate $B$
and defeat the sufficient condition, while the true-MDP drift again remains
positive (Table~\ref{tab:cycbottle}).

\begin{table}[t]
\centering
\small
\setlength{\tabcolsep}{4pt}
\begin{tabular}{@{}lrrrrr@{}}
\toprule
State & $V(s)$ & $\mathbb{E}_M[V']$ & $\mathbb{E}_{M'}[V']$ & Drift$(M)$ & Drift$(M')$ \\
\midrule
$s_0$ & $82.5$ & $81.225$ & $81.5$ & $1.275$ & $1$ \\
$s_1$ & $65$   & $63.942$ & $64$   & $1.058$ & $1$ \\
$s_2$ & $37.5$ & $36.717$ & $36.5$ & $0.783$ & $1$ \\
\bottomrule
\end{tabular}
\caption{Per-state certificate and drift values for the cyclic MDP with one
bottleneck forward probability $0.01$.}
\label{tab:cycbottle}
\end{table}

\subsection{Cyclic MDP, forward probabilities $0.1$}
\label{app:cyc01}
\begin{figure}[t]
\centering
\resizebox{\columnwidth}{!}{%
\begin{tikzpicture}[mdp,node distance=2cm]
\node[state] (s1) {$s_1$};
\node[state,right=of s1] (s2) {$s_2$};
\node[state,right=of s2] (s3) {$s_3$};
\node[state,accepting,right=of s3] (s4) {$s_4$};
\path[plab]
 (s1) edge[loop above] node {$0.7967$} (s1)
 (s2) edge[loop above] node {$0.8033$} (s2)
 (s3) edge[loop above] node {$0.8$} (s3)
 (s4) edge[loop above] node {$0.8$} (s4)
 (s1) edge[bend left] node {$0.1$} (s2)
 (s2) edge[bend left] node {$0.0967$} (s1)
 (s2) edge[bend left] node {$0.1$} (s3)
 (s3) edge[bend left] node {$0.1033$} (s2)
 (s3) edge[bend left] node {$0.0967$} (s4)
 (s4) edge[bend left] node {$0.1$} (s3)
 (s1) edge[bend right=25] node[below] {$0.1033$} (s4)
 (s4) edge[bend right=45] node[below] {$0.1$} (s1);
\end{tikzpicture}}
\caption{Cyclic MDP with forward probabilities $0.1$ (true kernel $M$).}
\label{fig:cyc01}
\end{figure}
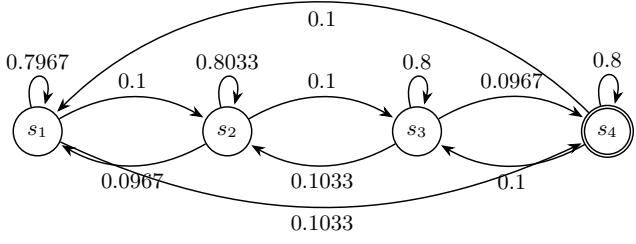

The third cyclic instance (Fig.~\ref{fig:cyc01}) raises \emph{all} forward
probabilities to $0.1$. With more frequent progress the hitting time, and hence
$B$, drops sharply, so that here the certificate \emph{provably} transfers
(Table~\ref{tab:cyc01}).

\begin{table}[t]
\centering
\small
\setlength{\tabcolsep}{4pt}
\begin{tabular}{@{}lrrrrr@{}}
\toprule
State & $V(s)$ & $\mathbb{E}_M[V']$ & $\mathbb{E}_{M'}[V']$ & Drift$(M)$ & Drift$(M')$ \\
\midrule
$s_0$ & $15$ & $13.95$  & $14$ & $1.05$  & $1$ \\
$s_1$ & $20$ & $19.017$ & $19$ & $0.983$ & $1$ \\
$s_2$ & $15$ & $14.067$ & $14$ & $0.933$ & $1$ \\
\bottomrule
\end{tabular}
\caption{Per-state certificate and drift values for the cyclic MDP with forward
probabilities $0.1$.}
\label{tab:cyc01}
\end{table}

\subsection{DAG-like MDP}
\label{app:dag}
\begin{figure}[t]
\centering
\resizebox{0.9\columnwidth}{!}{%
\begin{tikzpicture}[mdp]
\node[state] (s1) at (0,0) {$s_1$};
\node[state] (s2) at (2.6,1.0) {$s_2$};
\node[state] (s3) at (5.2,1.0) {$s_3$};
\node[state] (s4) at (2.6,-1.4) {$s_4$};
\node[state,accepting] (s5) at (5.2,-1.4) {$s_5$};
\path[plab]
 (s1) edge node {$0.5983$} (s2)
 (s1) edge[bend left=18] node {$0.4017$} (s3)
 (s2) edge node {$0.6017$} (s3)
 (s2) edge node[left] {$0.3983$} (s4)
 (s3) edge node[pos=0.35,above] {$0.6983$} (s4)
 (s3) edge node {$0.3017$} (s5)
 (s4) edge node {$1.0$} (s5)
 (s5) edge[loop below] node {$1.0$} (s5);
\end{tikzpicture}}
\caption{DAG-like MDP (true kernel $M$).}
\label{fig:dag}
\end{figure}
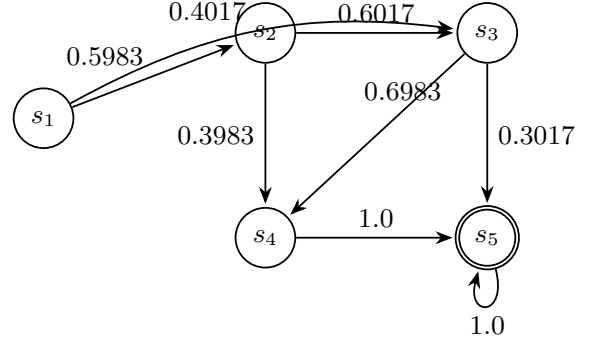

This benchmark (Fig.~\ref{fig:dag}) is an acyclic five-state MDP
$s_1,\dots,s_5$ with absorbing target $s_5$. Because every path makes steady
forward progress, $B$ stays small and the certificate transfers provably, with a
comfortable margin (Table~\ref{tab:dag}).

\begin{table}[t]
\centering
\small
\setlength{\tabcolsep}{4pt}
\begin{tabular}{@{}lrrrrr@{}}
\toprule
State & $V(s)$ & $\mathbb{E}_M[V']$ & $\mathbb{E}_{M'}[V']$ & Drift$(M)$ & Drift$(M')$ \\
\midrule
$s_0$ & $3.132$ & $2.131$ & $2.132$ & $1.001$ & $1$ \\
$s_1$ & $2.42$  & $1.421$ & $1.42$  & $0.999$ & $1$ \\
$s_2$ & $1.7$   & $0.698$ & $0.7$   & $1.002$ & $1$ \\
$s_3$ & $1$     & $0$     & $0$     & $1$     & $1$ \\
\bottomrule
\end{tabular}
\caption{Per-state certificate and drift values for the DAG-like MDP.}
\label{tab:dag}
\end{table}

\subsection{Gridworld MDP}
\label{app:grid}
\begin{figure}[t]
\centering
\resizebox{0.92\columnwidth}{!}{%
\begin{tikzpicture}[mdp]
\node[state] (s0) at (0,0)  {$s_0$};
\node[state] (s1) at (3,0)  {$s_1$};
\node[state] (s2) at (6,0)  {$s_2$};
\node[state] (s3) at (0,-3) {$s_3$};
\node[state] (s4) at (3,-3) {$s_4$};
\node[state] (s5) at (6,-3) {$s_5$};
\node[state] (s6) at (0,-6) {$s_6$};
\node[state] (s7) at (3,-6) {$s_7$};
\node[state,accepting] (s8) at (6,-6) {$s_8$};
\path[plab]
 (s0) edge[loop above] node {$0.0994$} (s0)
 (s1) edge[loop above] node {$0.0994$} (s1)
 (s8) edge[loop below] node {$1.0$} (s8)
 (s0) edge node {$0.8000$} (s1)
 (s1) edge[bend left] node {$0.8006$} (s2)
 (s2) edge[bend left] node {$0.1006$} (s1)
 (s3) edge node {$0.8006$} (s4)
 (s4) edge[bend left] node {$0.7994$} (s5)
 (s5) edge[bend left] node {$0.1006$} (s4)
 (s6) edge node {$0.8006$} (s7)
 (s7) edge node {$0.8006$} (s8)
 (s0) edge[bend left=15] node {$0.1006$} (s3)
 (s3) edge[bend left=15] node {$0.0994$} (s0)
 (s3) edge[bend left=15] node {$0.1000$} (s6)
 (s6) edge[bend left=15] node {$0.1000$} (s3)
 (s1) edge[bend left=15] node {$0.1000$} (s4)
 (s4) edge[bend left=15] node {$0.1000$} (s1)
 (s4) edge[bend left=15] node {$0.1006$} (s7)
 (s7) edge[bend left=15] node {$0.0994$} (s4)
 (s2) edge node[right] {$0.7994$} (s5)
 (s5) edge node[right] {$0.7994$} (s8);
\end{tikzpicture}}
\caption{Gridworld MDP (true kernel $M$; reproduced from the experiment output).}
\label{fig:grid}
\end{figure}
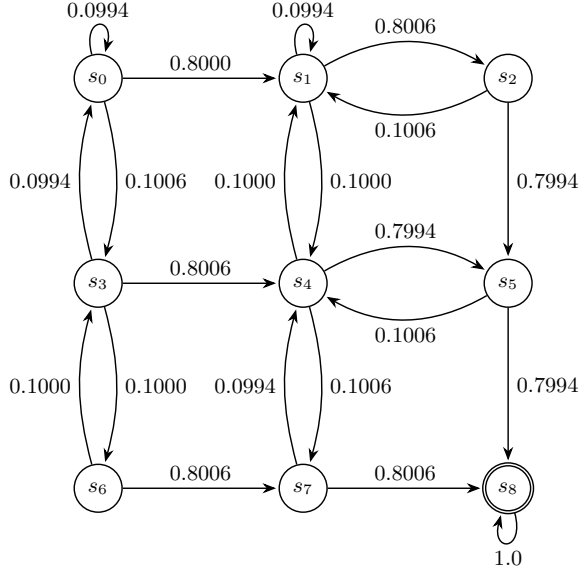

This benchmark (Fig.~\ref{fig:grid}) is a $3\times 3$ gridworld with states
$s_0,\dots,s_8$ and target $s_8$. Under the fixed policy each state advances
toward the goal (rightward or downward) with high probability and slips
laterally or self-loops with small probability. This progress-making structure
keeps $B$ small, so the certificate transfers provably (Table~\ref{tab:grid}).

\begin{table}[t]
\centering
\small
\setlength{\tabcolsep}{4pt}
\begin{tabular}{@{}lrrrrr@{}}
\toprule
State & $V(s)$ & $\mathbb{E}_M[V']$ & $\mathbb{E}_{M'}[V']$ & Drift$(M)$ & Drift$(M')$ \\
\midrule
$s_0$ & $5.001$ & $4.001$ & $4.001$ & $1.001$ & $1$ \\
$s_2$ & $2.793$ & $1.795$ & $1.793$ & $0.998$ & $1$ \\
$s_4$ & $2.654$ & $1.654$ & $1.654$ & $1$     & $1$ \\
$s_6$ & $2.795$ & $1.794$ & $1.795$ & $1.001$ & $1$ \\
$s_7$ & $1.406$ & $0.404$ & $0.406$ & $1.002$ & $1$ \\
\bottomrule
\end{tabular}
\caption{Per-state certificate and drift values for the gridworld MDP.}
\label{tab:grid}
\end{table}

\subsection{Chain (RiverSwim-style) MDP}
\label{app:chain}
\begin{figure}[t]
\centering
\resizebox{\columnwidth}{!}{%
\begin{tikzpicture}[mdp,node distance=1.7cm]
\node[state] (s1) {$s_1$};
\node[state,right=of s1] (s2) {$s_2$};
\node[state,right=of s2] (s3) {$s_3$};
\node[state,right=of s3] (s4) {$s_4$};
\node[state,right=of s4] (s5) {$s_5$};
\node[state,accepting,right=of s5] (s6) {$s_6$};
\path[plab]
 (s1) edge[loop above] node {$0.5990$} (s1)
 (s2) edge[loop above] node {$0.6010$} (s2)
 (s3) edge[loop above] node {$0.5990$} (s3)
 (s4) edge[loop above] node {$0.6000$} (s4)
 (s5) edge[loop above] node {$0.5990$} (s5)
 (s6) edge[loop above] node {$1.0$} (s6)
 (s1) edge[bend left] node {$0.4010$} (s2)
 (s2) edge[bend left] node {$0.0500$} (s1)
 (s2) edge[bend left] node {$0.3490$} (s3)
 (s3) edge[bend left] node {$0.0510$} (s2)
 (s3) edge[bend left] node {$0.3500$} (s4)
 (s4) edge[bend left] node {$0.0490$} (s3)
 (s4) edge[bend left] node {$0.3510$} (s5)
 (s5) edge[bend left] node {$0.0500$} (s4)
 (s5) edge[bend left] node {$0.3510$} (s6);
\end{tikzpicture}}
\caption{Chain (RiverSwim-style) MDP (true kernel $M$).}
\label{fig:chain}
\end{figure}
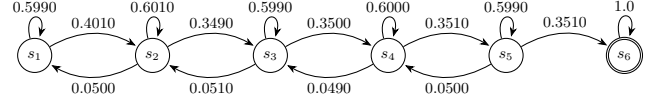

This benchmark (Fig.~\ref{fig:chain}) is a six-state RiverSwim-style chain
$s_1,\dots,s_6$ with target $s_6$: each interior state predominantly self-loops,
advances toward the target with moderate probability, and occasionally drifts
backward. The certificate transfers provably here (Table~\ref{tab:chain}).

\begin{table}[t]
\centering
\small
\setlength{\tabcolsep}{4pt}
\begin{tabular}{@{}lrrrrr@{}}
\toprule
State & $V(s)$ & $\mathbb{E}_M[V']$ & $\mathbb{E}_{M'}[V']$ & Drift$(M)$ & Drift$(M')$ \\
\midrule
$s_0$ & $15.695$ & $14.692$ & $14.695$ & $1.003$ & $1$ \\
$s_1$ & $13.195$ & $12.198$ & $12.195$ & $0.997$ & $1$ \\
$s_3$ & $6.664$  & $5.657$  & $5.664$  & $1.007$ & $1$ \\
$s_4$ & $3.333$  & $2.33$   & $2.333$  & $1.003$ & $1$ \\
\bottomrule
\end{tabular}
\caption{Per-state certificate and drift values for the chain (RiverSwim-style)
MDP.}
\label{tab:chain}
\end{table}

\subsection{Large $100$-state MDP}
\label{app:large}
Our final benchmark is a large, sparse MDP with $100$ states $s_0,\dots,s_{99}$;
being high-dimensional, it is reported by the representative per-state slice below
rather than a drawn automaton. Its structure contains a single rare transition
whose large hitting-time value dominates $B$, which is what makes the sufficient
condition fail by a wide margin while the true-MDP drift stays positive at every
state (Table~\ref{tab:large}).

\begin{table}[t]
\centering
\footnotesize
\setlength{\tabcolsep}{3.5pt}
\begin{tabular}{@{}lrrrrr@{}}
\toprule
State & $V(s)$ & $\mathbb{E}_M[V']$ & $\mathbb{E}_{M'}[V']$ & Drift$(M)$ & Drift$(M')$ \\
\midrule
$s_{0}$  & $421.467$    & $420.445$   & $420.467$   & $1.021$ & $1$ \\
$s_{25}$ & $8791.467$   & $8790.446$  & $8790.467$  & $1.021$ & $1$ \\
$s_{50}$ & $10661.467$  & $10660.467$ & $10660.467$ & $1$     & $1$ \\
$s_{75}$ & $8364.8$     & $8363.778$  & $8363.8$    & $1.022$ & $1$ \\
$s_{98}$ & $425.2$      & $424.179$   & $424.2$     & $1.021$ & $1$ \\
\bottomrule
\end{tabular}
\caption{Per-state certificate and drift values for the large $100$-state MDP
(representative states).}
\label{tab:large}
\end{table}

\section{Omitted Proofs}
\label{app:omitted}

This supplementary material collects the proofs omitted from the main body. Each proof is
restated together with the statement it establishes.

\subsection{Proofs from Section~\ref{sec:prelim}}

\begin{proof}[Proof of Theorem~\ref{thm:probability_one_if_rsm_exists}]
The strategy is to build a nonnegative supermartingale that would have to grow
without bound if the target were missed, and then invoke
Theorem~\ref{thm:supermartingale_inequality} to rule this out. Fix any
$s \in S$; the argument does not use that the chain starts at $s_0$, which is
what gives the conclusion at every state. Under $\pi$, $M$ induces a Markov
chain $(X_k)_{k \geq 0}$ with $X_0 = s$. Let
$\tau = \inf\{k \geq 0 \mid X_k \in T\}$ and set
$Y_k = V(X_{k \wedge \tau}) + \varepsilon\,(k \wedge \tau)$, noting that $\tau$
is a stopping time, so $\{\tau > k\} \in \mathcal{F}_k$. Then $\{Y_k\}$ is a
nonnegative supermartingale with $Y_0 = V(s)$. Nonnegativity is condition~(i)
together with $\varepsilon (k \wedge \tau) \ge 0$. For the supermartingale
property we split on $\mathcal{F}_k$: on $\{\tau \le k\}$ the process is frozen,
$X_{(k+1) \wedge \tau} = X_{k \wedge \tau}$ and $(k+1) \wedge \tau = k \wedge
\tau$, so $Y_{k+1} = Y_k$ pathwise and condition~(ii) is never invoked at a state
of $T$; on $\{\tau > k\}$ we have $X_k \in S \setminus T$, so condition~(ii)
gives $\mathbb{E}[V(X_{k+1}) \mid \mathcal{F}_k] \le V(X_k) - \varepsilon$ and
hence $\mathbb{E}[Y_{k+1} \mid \mathcal{F}_k] \le V(X_k) + \varepsilon k = Y_k$. Observe that on
$\{\tau = \infty\}$ we have $Y_k \geq \varepsilon k$, so
$\sup_k Y_k = \infty$. By Theorem~\ref{thm:supermartingale_inequality}, for every
$\lambda > 0$,
$\mathbb{P}^{\pi}_{s}[\sup_k Y_k \geq \lambda] \leq V(s)/\lambda$. Since
$\{\tau = \infty\} \subseteq \{\sup_k Y_k \geq \lambda\}$, letting
$\lambda \to \infty$ forces $\mathbb{P}^{\pi}_{s}[\tau = \infty] = 0$, i.e.\
$\mathbb{P}^{\pi}_{s}[\mathrm{Reach}(T)] = 1$.
\end{proof}

\begin{proof}[Proof of Corollary~\ref{cor:existence_almost_sure}]
The forward direction is Theorem~\ref{thm:probability_one_if_rsm_exists}, which
yields $\mathbb{P}^{\pi}_{s}[\mathrm{Reach}(T)] = 1$ at every $s$. For the
converse, assume $\mathbb{P}^{\pi}_{s}[\mathrm{Reach}(T)] = 1$ for every
$s \in S$, and let $\tau = \inf\{k \ge 0 \mid X_k \in T\}$ as above.

We first check that $\tau$ is \emph{integrable} from every state. Almost-sure
finiteness alone does not give this --- a random variable can be finite almost
surely and still have infinite mean --- and the step uses that $S$ is finite.
Write $n = |S|$ and let $q > 0$ be the smallest nonzero transition probability of
$M$ under $\pi$. Since $\mathbb{P}^{\pi}_{s}[\mathrm{Reach}(T)] = 1$, every state
has a path to $T$ in the transition graph induced by $\pi$, and a shortest such
path is simple, hence of length at most $n - 1$. Following it reaches $T$ within
$n - 1$ steps with probability at least $q^{\,n-1}$, from every state. Applying
this at the start of each block of $n - 1$ steps --- legitimate by the Markov
property, since the same bound holds from whatever state the chain occupies
there --- gives
\[
\mathbb{P}^{\pi}_{s}\bigl[\tau > k(n-1)\bigr] \;\le\; \bigl(1 - q^{\,n-1}\bigr)^{k},
\]
a geometric tail, so $\mathbb{E}_s[\tau] < \infty$ for every $s$.

Hence $V(s) = \mathbb{E}_s[\tau]$ is a well-defined map $S \to \mathbb{R}_{\ge 0}$.
It satisfies $V \geq 0$; $V \equiv 0$ on $T$, since $\tau = 0$ there under the
convention $\inf\{k \ge 0\}$; and, decomposing on the first step for
$s \in S \setminus T$ (using the Markov property of the chain induced by the
memoryless $\pi$),
$V(s) = 1 + \mathbb{E}_{s' \sim P(\cdot \mid s, \pi(s))}[V(s')]$. This is exactly
the $\varepsilon$-decrease condition with $\varepsilon = 1$, so $V$ is a
$1$-RSM.
\end{proof}

\subsection{Proofs from Section~\ref{sec:exp_hit}}

\begin{proof}[Proof of Lemma~\ref{lem:pmin_progress_path_sampling}]
The strategy is to bound the length of a shortest progress path, lower-bound the
probability of following it, and then amplify by independent episodes.

Let $n=|S|$. Since $T$ is reached almost surely under $\pi$ from every state,
every state has a path to $T$ in the transition graph induced by $\pi$. For a state
$s$, let $\operatorname{Dist}(s)$ be the length of a shortest such path, so
$\operatorname{Dist}(s)=0$ for $s\in T$ and, since the MDP is finite,
$\operatorname{Dist}(s)\le n-1$ for every reachable $s$. For every non-target
reachable state $s$ there is a successor $t$ with $P(t\mid s,\pi(s))>0$ and
$\operatorname{Dist}(t)=\operatorname{Dist}(s)-1$; we call such a transition a
\emph{progress transition}. By Definition~\ref{def:p_min}, each progress transition has
probability at least $p_{\min}$.

Let $D=\max_{s}\operatorname{Dist}(s)$ over reachable states, so $D\le n-1$.
Starting from any reachable state, there is a path of at most $D$ progress
transitions leading to $T$, and hence the probability of reaching $T$ within the
next $D$ steps is at least $p:=(p_{\min})^{D}$.

Now consider independent episodes of length $D$. In each episode the probability
of reaching $T$ is at least $p$, so the probability of not reaching $T$ in a
single episode is at most $1-p$, and after $m$ episodes it is at most $(1-p)^m$.
To drive this failure probability to at most $\delta$, it suffices to take
\[
\begin{aligned}
(1-p)^m&\;\le\;\delta,\\
\text{i.e.,}\quad m&\;\ge\;\frac{\log(\delta)}{\log(1-p)},
\end{aligned}
\]
where the direction of the inequality is preserved because $\log(1-p)<0$. Each
episode uses at most $D$ samples, so it suffices to take
\[
N\;=\;mD\;\ge\;D\cdot\frac{\log(\delta)}{\log\!\left(1-(p_{\min})^{D}\right)}.
\]
Finally, using the worst-case bound $D\le n-1$ (the expression is increasing in
$D$), it suffices to take
\[
N\;\ge\;(n-1)\cdot\frac{\log(\delta)}{\log\!\left(1-(p_{\min})^{n-1}\right)}.
\]
Thus, with probability at least $1-\delta$, a path to $T$ is observed within a
number of samples bounded by the known parameters $n$, $p_{\min}$, and $\delta$.
\end{proof}

\begin{proof}[Proof of Theorem~\ref{thm:valid_rsm_proof}]
The strategy is to verify the two defining conditions of
Definition~\ref{def:rsm_for_MDP} directly from the recursive form of
Definition~\ref{def:exp_hit_time}. Throughout, set $V(s) = \mathbb{E}_s^\pi[H_T]$.

\medskip
\noindent \textbf{(1) Non-negativity.}
For every $s \in S$, the hitting time $H_T$ is a non-negative random variable, and
hence
\[
V(s) = \mathbb{E}_s^\pi[H_T] \ge 0.
\]

\medskip
\noindent \textbf{(2) Expected decrease.}
Let $s \in S \setminus T$ and let $a = \pi(s)$. By Definition~\ref{def:exp_hit_time},
\[
V(s) = 1 + \sum_{y \in S} P(y \mid s, a)\, V(y),
\]
and the sum is exactly the one-step expectation of $V$,
\[
\mathbb{E}_{t \sim P(s,a)}[V(t)] = \sum_{y \in S} P(y \mid s, a)\, V(y).
\]
Hence $V(s) = \mathbb{E}_{t \sim P(s,a)}[V(t)] + 1$, which is precisely the
expected-decrease condition with $\varepsilon = 1$. Both conditions holding, $V$ is
a $1$-RSM.
\end{proof}

\begin{proof}[Proof of Theorem~\ref{thm:finite_hitting_time}]
The high-level idea is to show that from any state of finite distance the target
is reached within $D$ steps with probability at least $(p_{\min})^D$, and then to
turn this per-block success probability into a geometric bound on the expected
hitting time.

By hypothesis $\mathrm{Dist}(x) < \infty$ for every $x \in S$. Fix a state
$s \in S$. If $s \in T$, then $H_T = 0$ and hence $\mathbb{E}_s^\pi[H_T] = 0$, so
assume $s \notin T$. By Definition~\ref{def:dist_to_target}, every state
$x \in S \setminus T$ admits a successor $t \in \mathrm{Post}(x, \pi(x))$
with $\mathrm{Dist}(t) = \mathrm{Dist}(x) - 1$. By
Definition~\ref{def:progress_making}, the transition $x \to t$ is progress-making, and
since $t \in \mathrm{Post}(x, \pi(x))$ we have $P(t \mid x, \pi(x)) > 0$;
therefore, by the definition of $p_{\min}$,
\[
P(t \mid x, \pi(x)) \ge p_{\min}.
\]

Now consider any state $x \in S$. Repeatedly following
a progress-making transition decreases the distance by one at each step, so
starting from $x$ there is a path to $T$ of length
$\mathrm{Dist}(x) \le D$. Since each transition along this path has probability at
least $p_{\min}$, the probability of following it and reaching $T$ within $D$
steps is at least $(p_{\min})^D$. Hence, for every state $x \in S$,
\[
\mathbb{P}_x^\pi[H_T \le D] \ge (p_{\min})^D.
\]

Let $p = (p_{\min})^D > 0$, and divide time into blocks of length $D$. If the
target has not been reached before the start of a block, the process occupies
some state $x$, and $\mathrm{Dist}(x) < \infty$ by hypothesis --- this is exactly
where reachability \emph{from every state} is needed, since $\{\mathrm{Dist} <
\infty\}$ is not in general closed under transitions --- and hence reaches $T$
during that block with probability at least $p$; equivalently, it fails to reach $T$ during
the block with probability at most $1 - p$. It follows that, for every
$k \ge 0$,
\[
\mathbb{P}_s^\pi[H_T > kD] \le (1 - p)^k.
\]
Since $\mathbb{E}_s^\pi[H_T] = \sum_{t \ge 0} \mathbb{P}_s^\pi[H_T > t]$, grouping
the sum into blocks of length $D$ gives
\[
\begin{aligned}
\mathbb{E}_s^\pi[H_T]
&\le \sum_{k=0}^{\infty} D \cdot \mathbb{P}_s^\pi[H_T > kD] \\
&\le D \sum_{k=0}^{\infty} (1 - p)^k
= D \cdot \frac{1}{p} = \frac{D}{(p_{\min})^D}.
\end{aligned}
\]
In particular, $\mathbb{E}_s^\pi[H_T]$ is finite.
\end{proof}

\begin{proof}[Proof of Theorem~\ref{thm:transferability_under_pmin}]
This is the specialization of Theorem~\ref{thm:general_certificate_transferability} to
the expected hitting-time certificate, for which $\varepsilon' = 1$; we spell it
out for completeness. Fix $x \in S$ and let $a = \pi(x)$. By definition of
expectation on the two models,
\[
\begin{aligned}
\mathbb{E}_M[V(X_{t+1}) \mid x, a] &= \sum_{y} P(y \mid x, a)\, V(y), \\
\mathbb{E}_{M'}[V(X_{t+1}) \mid x, a] &= \sum_{y} P'(y \mid x, a)\, V(y).
\end{aligned}
\]
Subtracting and taking absolute values, so as to invoke the approximation
assumption,
\[
\begin{aligned}
&\mathbb{E}_M[V(X_{t+1}) \mid x, a] - \mathbb{E}_{M'}[V(X_{t+1}) \mid x, a] \\
&\quad \le \sum_{y} |P(y \mid x, a) - P'(y \mid x, a)| \cdot |V(y)|.
\end{aligned}
\]
By Definition~\ref{def:total_prob_error}, $\sum_{y} |P(y \mid x, a) - P'(y \mid x, a)| \le
\eta$ for the fixed pair $(x, a)$, and since $V(s) = \mathbb{E}_s^\pi[H_T]$ is
finite for every $s$ (the progress paths of $M$ survive into $M'$ with edge
weight $p_{\min}/2$, as shown below, so $T$ is reachable from every state of $M'$
as well), we may set
$B = \max_{s \in S} V(s)$, so that $|V(y)| \le B$ for all $y$. Therefore,
\[
\mathbb{E}_M[V(X_{t+1}) \mid x, a]
\le \mathbb{E}_{M'}[V(X_{t+1}) \mid x, a] + B\eta.
\]
Because $V$ is a $1$-RSM on $M'$, we have
$\mathbb{E}_{M'}[V(X_{t+1}) \mid x, a] \le V(x) - 1$. Substituting,
\[
\mathbb{E}_M[V(X_{t+1}) \mid x, a]
\le V(x) - 1 + B\eta
= V(x) - (1 - B\eta).
\]
Writing $\varepsilon := 1 - B\eta$, the RSM condition on $M$ holds whenever
$\varepsilon > 0$, i.e., whenever $\eta < \tfrac{1}{B}$. It remains to bound $B$
a priori. Theorem~\ref{thm:finite_hitting_time} cannot be applied to $M'$ with
$p_{\min}$, since $p_{\min}$ constrains $P$ and not $P'$; an estimated transition
may carry arbitrarily small mass. Suppose therefore that
$\eta \le p_{\min}/2$. Every transition with $P(y \mid x, a) \ge p_{\min}$ then
satisfies $P'(y \mid x, a) \ge p_{\min} - \eta \ge p_{\min}/2$, because
Definition~\ref{def:total_prob_error} bounds each entry by $\eta$. Hence every
progress path of $M$ (Definition~\ref{def:progress_making}) is a path of $M'$ all
of whose edges have probability at least $p_{\min}/2$, and
Theorem~\ref{thm:finite_hitting_time} applied to $M'$ with $p_{\min}/2$ gives
$B \le \tfrac{D}{(p_{\min}/2)^D}$. The transfer condition is therefore implied by
$\eta < \tfrac{(p_{\min}/2)^D}{D}$, which for $D \ge 1$ is itself at most
$p_{\min}/2$, so the supposition above is not an extra hypothesis. Hence the
expected hitting-time function, an RSM on $M'$, is also an RSM on $M$.
\end{proof}

\begin{remark}
It is worth noting where the $p_{\min}$ assumption is actually spent. In the proof
above it enters only through the bound $B \le D/(p_{\min}/2)^D$ on the
certificate, which is what converts the a-posteriori transfer condition
$\eta < 1/B$ into the a-priori threshold $\eta < (p_{\min}/2)^D/D$. The halved
base is what makes the bound legitimate on the \emph{learned} model: $p_{\min}$
itself constrains only $P$. The transfer mechanism itself
(Theorem~\ref{thm:general_certificate_transferability}) needs no such assumption; it
requires only that the certificate be bounded.
\end{remark}

\section{The General Certification-Learning Framework}
\label{app:general_learning}

Section~\ref{sec:pac_imp} works with the specialization of the
certification-learning framework to almost-sure reachability. For completeness we
record here the general formulation, which applies to arbitrary specifications.

Formally, a \emph{certification learning task} is a tuple $(M,\pi,\varphi)$, where
$M$ is an unknown MDP with sampling access, $\pi$ is a policy in $M$, and $\varphi$
is a specification over $M$. A learning algorithm $\A$ is an iterative process that,
in each iteration, (i) either resets the simulator or takes a step in $M$, and (ii)
outputs its current estimate of a certificate $C_i$ for the policy $\pi$. Thus $\A$
induces a random sequence $\{C_i\}_{i=1}^{\infty}$ of output certificates, and
interacts with the kernel only through samples.

\begin{definition}[Learning Certificates with Probabilistic Guarantees]
\label{def:probabilisticguarante}
A learning algorithm $\A$ is said to \emph{learn a certificate with probabilistic
guarantees} for a class of specifications $\C$ if there exists a
\emph{sample-complexity function} $h$ such that, for all $\delta>0$ and all
certificate learning tasks $(M,\pi,\varphi)$ with
$M=(S,A,P,s_0)$ and $\varphi\in\C$, taking
\(
N=h\!\left(|S|,|A|,|\pi|,|\varphi|,\tfrac{1}{\delta}\right),
\)
we have, for all $n>N$,
\[
\Pr\!\left[
C_n \text{ certifies } \pi \text{ w.r.t. } \varphi
\right]
\;\geq\; 1-\delta.
\]
We say that a learning algorithm with probabilistic guarantees is
\emph{efficient} if the sample-complexity function $h$ is polynomial in
$|S|$, $|A|$, $|\pi|$, $|\varphi|$, and $\frac{1}{\delta}$.
\end{definition}

Here $|S|$ and $|A|$ denote the number of states and actions of the MDP, $|\pi|$
denotes the size of the representation of the policy $\pi$, and $|\varphi|$ denotes
the size of the specification $\varphi$. The crux of the definition is that it
demands a single budget $N$ that works \emph{uniformly} over all tasks of a given
size. Once $N$ is exceeded, correctness must hold with probability at least
$1-\delta$.

Almost-sure reachability is the instance in which $\varphi$ is a target set
$T\subseteq S$, and certifying $\pi$ with respect to $\varphi$ means establishing
$\mathbb{P}^{\pi}_{s_0}[\mathrm{Reach}(T)]=1$. Writing such tasks as $(M,\pi,T)$ and
absorbing $|T|$ and $|\pi|$ into the finite-MDP description size recovers
Definition~\ref{def:learning_rsm_prob_guarantee}, the reachability formulation used
in the main text.

\section{Ranking Supermartingales for Stochastic Systems}
\label{app:rsm_general}

Ranking supermartingales were introduced for general discrete-time stochastic
systems~\cite{lechner2022stability} $x_{k+1} = f(x_k, \pi(x_k), \omega_k)$, with
state $x_k \in X$, fixed controller $\pi$, and noise $\omega_k \sim d$. The MDP
certificate of Definition~\ref{def:rsm_for_MDP} is the instance of this notion in
which $X = S$ and the update is realized by sampling
$s' \sim P(\cdot \mid s, \pi(s))$.

\begin{definition}[RSM for a stochastic system~\cite{lechner2022stability}]
\label{def:rsm_general}
Let $\mathcal{X}_t \subseteq X$ be a target set. A continuous function
$V : X \to \mathbb{R}$ is a \emph{ranking supermartingale (RSM)} w.r.t.\
$\mathcal{X}_t$ if there exists $\varepsilon > 0$ such that
\begin{itemize}
    \item $V(x) \geq 0$ for all $x \in X$, and
    \item
$V(x) \geq \mathbb{E}_{\omega \sim d}[V(f(x, \pi(x), \omega))] + \varepsilon$
for all $x \in X \setminus \mathcal{X}_t$.
\end{itemize}
\end{definition}
\begin{theorem}[Almost-sure reachability~\cite{lechner2022stability}]
\label{thm:almost-sure}
If an RSM w.r.t.\ $\mathcal{X}_t$ exists, then
$\mathbb{P}_{x_0}[\mathrm{Reach}(\mathcal{X}_t)] = 1$ for every $x_0 \in X$.
\end{theorem}

\end{document}